\documentclass[twocolumn]{aa}  

\usepackage{graphicx}

\usepackage{txfonts}
\usepackage{gensymb,subfig}
\usepackage{booktabs}
\usepackage{threeparttable}

\usepackage{hyperref}  

\usepackage{orcidlink}
\newcommand\orcidicon[1]{\orcidlink{#1}}

\def\ts     {\thinspace} 
\usepackage{amsmath}

\def\kms  {\ifmmode{{\rm \ts km\ts s}^{-1}}\else{\ts km\ts s$^{-1}$\ts}\fi}
\def\msol {\ifmmode{{\rm M}_{\odot}}\else{M$_{\odot}$\ts}\fi}
\def\lsun {\ifmmode{{\rm L}_{\odot}}\else{L$_{\odot}$\ts}\fi}
\def\cii  {\ifmmode{{\rm [C}{\rm \small II}]}\else{[C\ts {\scriptsize II}]\ts}\fi}
\def\ci   {\ifmmode{{\rm C}{\rm \small I}}\else{C\ts {\scriptsize I}\ts}\fi}
\def\m    {\ifmmode{\mu {\rm m}}\else{$\mu$m}\fi}
\def\hi   {\ifmmode{{\rm H}{\rm \small I}}\else{H\ts {\scriptsize I}\ts}\fi}
\def\hii  {\ifmmode{{\rm H}{\rm \small II}}\else{H\ts {\scriptsize II}\ts}\fi}
\def\nii  {\ifmmode{{\rm [N}{\rm \small II}]}\else{[N\ts {\scriptsize II}]\ts}\fi}
\def\oiii {\ifmmode{{\rm [O}{\rm \small III}]}\else{[O\ts {\scriptsize III}]\ts}\fi}
\def\hh   {\ifmmode{{\rm H}_2}\else{H$_2$\ts}\fi}
\def\nhh  {\ifmmode{N({\rm H}_2)}\else{$N$(H$_2$)\ts}\fi}
\def\microns {\ifmmode{\mu{\rm m}}\else{$\mu$m\ts}\fi}

\begin{document} 

\title{BASS. XLIV. Morphological preferences of local hard X-ray selected AGN}

\author{
Miguel Parra Tello\inst{1\orcidicon{0000-0001-5649-7798}}
\and
Franz E. Bauer\inst{2\orcidicon{0000-0002-8686-8737}}
\and
Demetra De Cicco\inst{3,4,5\orcidicon{0000-0001-7208-5101}}
\and
Goran Doll\inst{3,4}
\and
Michael Koss\inst{6}
\and
Ezequiel Treister\inst{2}
\and
Carolina Finlez\inst{1}
\and
Marco Troncoso\inst{1}
\and
Connor Auge\inst{6}
\and
I. del Moral-Castro\inst{1\orcidicon{0000-0001-8931-1152}}
\and
Aeree Chung\inst{7\orcidicon{0000-0003-1440-8552}}
\and 
Kriti K. Gupta\inst{9,10\orcidicon{0009-0007-9018-1077}}
\and 
Jeein Kim\inst{7\orcidicon{0000-0002-3170-7434}}
\and 
Kyuseok Oh\inst{8\orcidicon{0000-0002-5037-951X}}
\and
Claudio Ricci\inst{12, 13\orcidicon{0000-0001-5231-2645}}
\and 
Federica Ricci\inst{14,15\orcidicon{0000-0001-5742-5980}}
\and 
Alejandra Rojas\inst{11\orcidicon{0000-0003-0006-8681}}
\and
Turgay Caglar\inst{16,17\orcidicon{0000-0002-9144-2255}}
\and
Fiona Harrison\inst{18\orcidicon{0000-0002-4226-8959}}
\and
Meredith C. Powell\inst{19\orcidicon{0000-0003-2284-8603}}
\and 
Daniel Stern\inst{20\orcidicon{0000-0003-2686-9241}}
\and
Benny Trakhtenbrot\inst{21\orcidicon{0000-0002-3683-7297}}
\and 
C. Megan Urry\inst{22,23\orcidicon{0000-0002-0745-9792}}
}

\institute{Instituto de Astrof\'isica, Facultad de F\'isica, Pontificia Universidad Cat\'olica de Chile, Casilla 306, Santiago 22, Chile\\
              \email{mgparra@uc.cl}
         \and
{Instituto de Alta Investigaci{\'{o}}n, Universidad de Tarapac{\'{a}}, Casilla 7D, Arica, Chile} 
         \and
{Dipartimento di Fisica "Ettore Pancini", Università di Napoli Federico II, via Cinthia 9, 80126 Napoli, Italy}
        \and 
{INAF - Osservatorio Astronomico di Capodimonte, Salita Moiariello 16, 80131 Napoli, Italy}
        \and
{Millennium Institute of Astrophysics, MAS,  Monse\~{n}or Nuncio S\'{o}tero Sanz 100, Providencia, Santiago de Chile.} 
         \and
{Eureka Scientific, 2452 Delmer Street Suite 100, Oakland, CA 94602-3017, USA}
         \and
{Department of Astronomy, Yonsei University, 50 Yonsei-ro, Seodaemun-gu, Seoul 03722, Korea }
        \and 
{Korea Astronomy and Space Science Institute, Daedeokdae-ro 776, Yuseong-gu, Daejeon 34055, Republic of Korea}
        \and 
{STAR Institute, Li\`ege Universit\'e, Quartier Agora - All\'ee du six Ao\^ut, 19c B-4000 Li\`ege, Belgium}
        \and 
{Sterrenkundig Observatorium, Universiteit Gent, Krijgslaan 281 S9, B-9000 Gent, Belgium}
        \and
{Departamento de F\'isica, Universidad T\'ecnica Federico Santa Mar\'ia, Vicu\~{n}a Mackenna 3939, San Joaqu\'in, Santiago de Chile, Chile}
        \and 
{Instituto de Estudios Astrofísicos, Facultad de Ingeniería y Ciencias, Universidad Diego Portales, Av. Ejército Libertador 441, Santiago, Chile}
        \and 
{Kavli Institute for Astronomy and Astrophysics, Peking University, Beijing 100871, People's Republic of China}
        \and 
{Dipartimento di Matematica e Fisica, Universit`a degli Studi Roma Tre, Via della Vasca Navale 84, I-00146, Roma, Italy}
        \and 
{INAF - Osservatorio Astronomico di Roma, via Frascati 33, 00040 Monteporzio Catone, Italy}
        \and 
{George P. and Cynthia Woods Mitchell Institute for Fundamental Physics and Astronomy, Texas A\&M University, College Station, TX, 77845, USA}
        \and 
{Leiden Observatory, PO Box 9513, 2300 RA Leiden, The Netherlands}
        \and
{Cahill Center for Astronomy and Astrophysics, California Institute of Technology, Pasadena, CA 91125, USA}
        \and 
{Leibniz-Institut f\"ur Astrophysik Potsdam (AIP), An der Sternwarte 16, 14482 Potsdam, Germany}
        \and 
{Jet Propulsion Laboratory, California Institute of Technology, 4800 Oak Grove Drive, MS 169-224, Pasadena, CA 91109, USA}
        \and 
{School of Physics and Astronomy, Tel Aviv University, Tel Aviv 69978, Israel}
        \and 
{Yale Center for Astronomy \& Astrophysics and Department of Physics, Yale University, P.O. Box 208120, New Haven, CT 06520-8120, USA}
        \and 
{Department of Physics, Yale University, P.O. Box 208120, New Haven, CT 06520, USA}
}

    \date{Submitted 1 April 2025 / Revised 27 May 2025 / Accepted 12 June 2025}

\abstract 
  {We present detailed morphological classifications for the host of 1189 hard-X-ray selected (14--195\,keV) active galactic nuclei (AGNs) from the \textit{Swift}-BAT 105-month catalog as part of the BAT AGN Spectroscopic Survey (BASS). BASS provides a powerful all-sky census of nearby AGN, minimizing obscuration biases and providing a robust dataset for studying AGN-host galaxy connections.
  Classifications are based on a volunteer-based visual inspection workflow on the Zooniverse platform, adapted from the Galaxy Zoo DECaLS (GZD) project. Dual-contrast \textit{grz} color composite images, generated from public surveys (e.g., NOAO Legacy Survey, Pan-STARRS, SDSS) and dedicated observations enabled key morphological features to be identified.
Our analysis reveals that, with respect to a control sample of inactive galaxies, BASS AGN hosts show a deficiency of smooth elliptical galaxies ($\sim$70\%) and spiral galaxies with prominent arms ($\sim$80\%), while displaying an excess of mergers or disturbed systems ($\sim$400\%), and disk galaxies without a spiral structure ($\sim$300\%). These trends are found after controlling for redshift and \textit{i}-band magnitude, which suggests a preference for AGN activity in gas-rich, dynamically disturbed environments or transitional disk systems. We also find a higher bar fraction among AGN hosts than in a control sample ($\sim$50\% vs. $\sim$30\%).
We further explore the relationships between AGN properties (e.g., X-ray luminosity, black hole mass, and Eddington ratio) and host morphology, and find that high-luminosity and high-accretion AGN preferentially reside in smooth or point-like hosts. At the same time, lower-luminosity AGN are more common in disk galaxies. These results underscore the importance of morphological studies in understanding the fueling and feedback mechanisms that drive AGN activity and their role in galaxy evolution. Our dataset provides a valuable benchmark for future multiwavelength surveys (e.g. LSST, Roman, and Euclid) and automated morphological classification efforts.
}


   \maketitle
\section{Introduction}

The morphological properties of galaxies 
are shaped by secular processes and external interactions, and studying them provides clues to their formation and evolutionary history \citep[e.g.,][]{Hubble1926, Kennicutt1998_SFRHubbleReview, Blanton2009_GalaxyReview, Conselice2014_MorphologyReview, Bland-Hawthorn2016_MW_Review}. The host galaxies of active galactic nuclei (AGNs), in particular, offer an important physical context for understanding the mechanisms that drive the coeval growth of supermassive black holes (SMBHs) and spheroid stellar mass, the potential role of related feedback processes, and the links to the broader extragalactic environment \citep[e.g.,][]{Cisternas2011_No_Major_Mergers,  Woodrum2024_agn_in_green_valley, Dubois2016_agn_feedback_and_morphologies, Man2019_environment_driver}. 
During their AGN phases, SMBHs have the potential to drive the evolution of the host galaxy via radiation, winds, and jets \citep[e.g.,][]{DiMatteo2005_Feedback_sim, Hopkins2006_Merger_Model}; such feedback mechanisms are capable of quenching their hosts \citep{Kormendy2013} by preventing gas from cooling and forming stars.\footnote{Although theoretical studies often rely on AGN feedback to produce realistic results \citep[e.g.,][]{Dubois2016_agn_feedback_and_morphologies}, direct observational evidence remains relatively elusive and/or divisive \citep[e.g.,][]{Scholtz2018_agn_feedback_and_sfr, Koss2021_BASS_Molgas}.} Thus, exploring the relationship between the nucleus and the state of its host galaxy is of special importance for galactic evolution.

Historically, the relationship between AGN activity and galaxy morphology has been a topic of significant debate. Early studies suggested a dichotomy between AGN types, with Seyfert galaxies predominantly residing in disk-dominated hosts \citep[e.g.,][]{Weedman1977_Seyfert_Review,Adams1977_Seyfert_Spirals}, while radio-loud AGNs were preferentially associated with massive ellipticals \citep[e.g.,][]{Matthews1964_RLAGN_ELL,Kauffmann2003_Morph_early,Best2005_Host_RLAGN}. With the advent of large-scale multiwavelength surveys, this picture has become more nuanced, finding that AGN hosts span a wide range of morphologies, with evidence for both merger-induced and secular fueling mechanisms \cite[e.g.,][]{Koss2011_BASSMorphs, Ricci2017_Merger_growth, Ellison2019_AGN_Merger_connect, Zhuang2023_AGN_Evolution}. The latter include bar structures, which are not only known to funnel gas to the inner regions of the galaxy \citep[e.g.,][]{shlosman1989_bars_in_bars, friedli1993_bar_gas_funnel}, but are also suspected of having a higher incidence among AGN hosts \citep[e.g.,][]{laine2002_more_bars_sy} and to positively correlate with higher accretion rates \citep[e.g.,][]{alonso2018_bars_and_accrats}. That said, there is extensive and conflicting observational literature on bars and AGNs that advocates for both independence \citep[e.g.,][]{cisternas2013_no_link_bar_fuel, goulding2017_bars_no_influence} or correlation \citep[e.g.,][]{oh2012_bar_effects, garland2024_bar_strength_agn}.

The masses of SMBHs have been intimately linked to their host spheroid masses \citep[e.g.,][]{Kormendy2013}, which reflects a fundamental connection between SMBH growth and galaxy evolution \citep[e.g.,][]{caglar2020_msigma, caglar2023_msigma-bass}. These spheroid masses comprise a key component of host morphologies and highlight the intertwined nature of structural and dynamical properties in galaxies. Yet within this SMBH-spheroid framework, it remains unclear how other critical variables such as accretion rate, AGN orientation, and the galaxy's merger history relate to the current morphology of AGN hosts. For instance, do host morphologies and structural features differ systematically across AGN luminosity or accretion state regimes? It is also not clear how major and minor mergers, bar structures, or disk instabilities contribute to the onset and fueling of AGN activity. 
Finally, it is not well understood how well-known AGN selection effects may influence the resulting morphological characterization of selected samples.
Addressing these questions requires comprehensive analyses of statistically significant AGN samples, particularly at low redshifts ($z \lesssim 0.15$), where the disentanglement of host galaxy features from AGN light is most feasible. Such studies are essential to uncover the interplay between AGN activity and galactic properties across diverse environments and evolutionary stages.

Here we aim to contribute to this ongoing discussion by examining the morphologies of nearby AGN host galaxies using a combination of high-quality imaging and robust morphological classification methods. To obtain a better census of AGN, and get a clearer picture of how they relate to their host galaxies, we need to construct samples that are relatively unbiased toward obscured AGN. Hard X-ray emission ($>$ 10 keV) provides a relatively robust selection criterion in this sense \citep[e.g.,][]{Ricci2017_BAT_X-ray}. 
Notably, the Burst Alert Telescope \citep[BAT][]{swift_bat} on board of the \textit{Swift} observatory has uniformly observed the entire sky in the 14--195 keV band, and detected a wide variety of hard X-ray sources, including many AGNs. A selection such as this one is less prone to the effects of AGN obscuration \citep{hickox&alexander_obscured_agn_review}, because the highly energetic nature of these photons allows them to pass through large amounts of dust and gas, which enables the identification of heavily obscured AGNs even at Compton-thick densities  \citep[$\log N_\mathrm{H} > 24$;][]{ricci2015_compton, KOss2016_swift_compton}. Robust classifications of AGNs among the \textit{Swift}-BAT-detected sources have been carried out as part of the BAT AGN Spectroscopic Survey\footnote{\url{https://bass-survey.com}} \citep[BASS;][]{koss2017_bas70s, Koss2022_BASSDR2_overview} across the entire sky and regardless of obscuration. A previous study by \cite{koss2010_mergers} set a precedent for a higher incidence of mergers and companions on BAT AGNs, both with respect to inactive galaxies and optically selected AGNs. These results were later supported by \cite{cotini2013_bass_mergers}, who also reported a higher merger fraction in the active population of BAT AGNs when compared to inactive galaxies.

In this work, we carry out a visual morphological classification and analysis of 1189 AGN host galaxies in BASS, which involves classifying these host galaxies into various categories (such as smooth, disks with and without arms, edge-on disks, and mergers) and comparing these morphologies to a control sample of already classified galaxies, matched in redshift and $i$-band magnitude. In Sect. \ref{sec:data}, we describe the sample, types of images used, and image processing, while in Sect. \ref{sec:zooniverse} we describe the volunteer visual classification scheme. The control sample is detailed in Sect. \ref{sec:gzd}. In Sect. \ref{sec:classifications} and Sect. \ref{sec:reclassifications}, we describe how volunteer responses are converted into broad morphological classes. We then present the results for the BASS sample by itself in Sect. \ref{sec:results_BASS} and with respect to the control sample in Sect. \ref{sec:results_colormag} and Sect. \ref{sec:results_comp}, while conclusions and discussion of future prospects can be found in Sect. \ref{sec:conclusions}. Throughout, we assume a flat $\Lambda$CDM cosmology with $H_0 = 69.32$ km/s/Mpc and $\Omega_m = 0.24$ \citep{WMAP9}.

\section{Data}\label{sec:data}

\subsection{BASS sample}\label{sec:data_bass}

We began with the \textit{Swift}-BAT 105-month catalog, which contains 1632 objects selected in the 14--195\,keV X-ray band across the entire sky, of which 1105 were listed as classified AGN in \citet{Oh2018_BAT105month}; an additional 84 previously unclassified sources have since been classified as AGN sources by BASS (M. Koss et al. in prep). We opt for this larger sample over the 70-month sample covered by BASS DR2, with the idea that the morphologies here will remain useful for both the DR2 and future DR3 samples. This parent AGN sample includes a variety of subtypes such as Seyfert 1--2s, LINERs, beamed, and previously ``unclassified'' AGN candidates. Stemming from the selection in the hard X-rays, the sample is less biased against obscured AGN, as reflected by the relative $\sim$50:50 balance between Seyfert 1 and 2 AGN \citep[e.g.][]{Oh2018_BAT105month, mejia-restrepo2022_sy, Koss2022_BASSDR2_overview}. 

\subsection{Image cutouts}\label{sec:data_cutouts}

To maximize image quality among the objects in our sample across several filters (in terms of sensitivity and spatial resolution), while minimizing the amount of different optical imaging sources (sky coverage), we constructed a hierarchical list of optical surveys to generate color image cutouts. In descending priority order, the surveys from which we used cutouts were Legacy Survey (DR10, hereafter LSDR10; \textit{grz} + partial \textit{i}; 0\farcs262/pixel; \citealt{decals_overview}), Pan-STARRS (DR2; \textit{grizy}; 0\farcs258/pixel; \citealt{panstarrs1dr2_overview, panstarrs1dr2_database}), SDSS (DR12; \textit{ugriz}; 0\farcs396/pixel; \citealt{sdssdr12_overview}) and DECaPS (DR2; \textit{grizY}; 0\farcs262/pixel; \citealt{decaps2_overview}). We adopted the \textit{grz} bands to generate RGB images, as these are the most commonly available among the surveys we draw from. 

A series of decisions and quality cuts for the individual band cutouts were adopted. We first must determine a reasonable size for the cutout, which should contain not only the full apparent extent of the host galaxy above the background, but extend by some factor beyond this to provide environmental context (are there any tidal features? companions?). To this end, we required as a minimum that the cutouts be 500$\times$500 pixels ($\approx$2\farcm2$\times$2\farcm2), while as a maximum, we adopted a physical extent of 60$\times$60 kpc$^{2}$ based on the angular distance from available redshift measurements. This worked well for most objects, allowing potentially relevant nearby companions to appear in the imaged region. A small number of problematic cases (7\%), where the images appeared too zoomed in or out, were identified based on visual inspection and corrected accordingly.\footnote{We considered scaling the size by the Petrosian radii, but discarded this possibility since we lacked this measurement for most objects and it is not well-constrained for major mergers.}

For the vast majority (1124/1189, or 94.5\%) of sources, all three \textit{grz} exposures from a given survey were available and considered usable. We required that for each \textit{g}, \textit{r}, or \textit{z}-band image, the target itself remained centered, the aforementioned extent was fully covered, and there were no major imaging artifacts (e.g., saturated stars, reflections, ghosts) or exposure gaps impairing assessment. For rejected images (seven cases; ``LS-archival hybrid''), we iterated through the surveys while looking for image cutouts of the objects in the three aforementioned optical bands. If an object was not imaged in the highest-priority survey or did not meet quality imaging standards, we searched for it in the next survey. We repeat this process either until a survey contains all three bands necessary for a color-composite image or until we fail to find them after exhausting the chain. In the latter scenarios (65 cases), exclusively in the Southern Hemisphere ($\delta {\le}{-}30\degree$) and typically outside of the combined footprint of the surveys, we found and incorporated archival \textit{griz} band imaging either from LSDR10 stacks (2 composed of \textit{gri} and 5 \textit{grz}) or DECam individual exposures (37 cases) using the NOIRlab datalab image-cutout archive;\footnote{\url{https://datalab.noirlab.edu/sia.php}} in 8 cases, we substituted useful \textit{i}-band imaging for missing \textit{z}-band images. Finally, for 21 galaxies that still lacked one or more cutouts, we acquired observations from the GROND and Sinistro instruments as detailed below. 

\begin{figure}
    \centering
    \includegraphics[width = \linewidth]{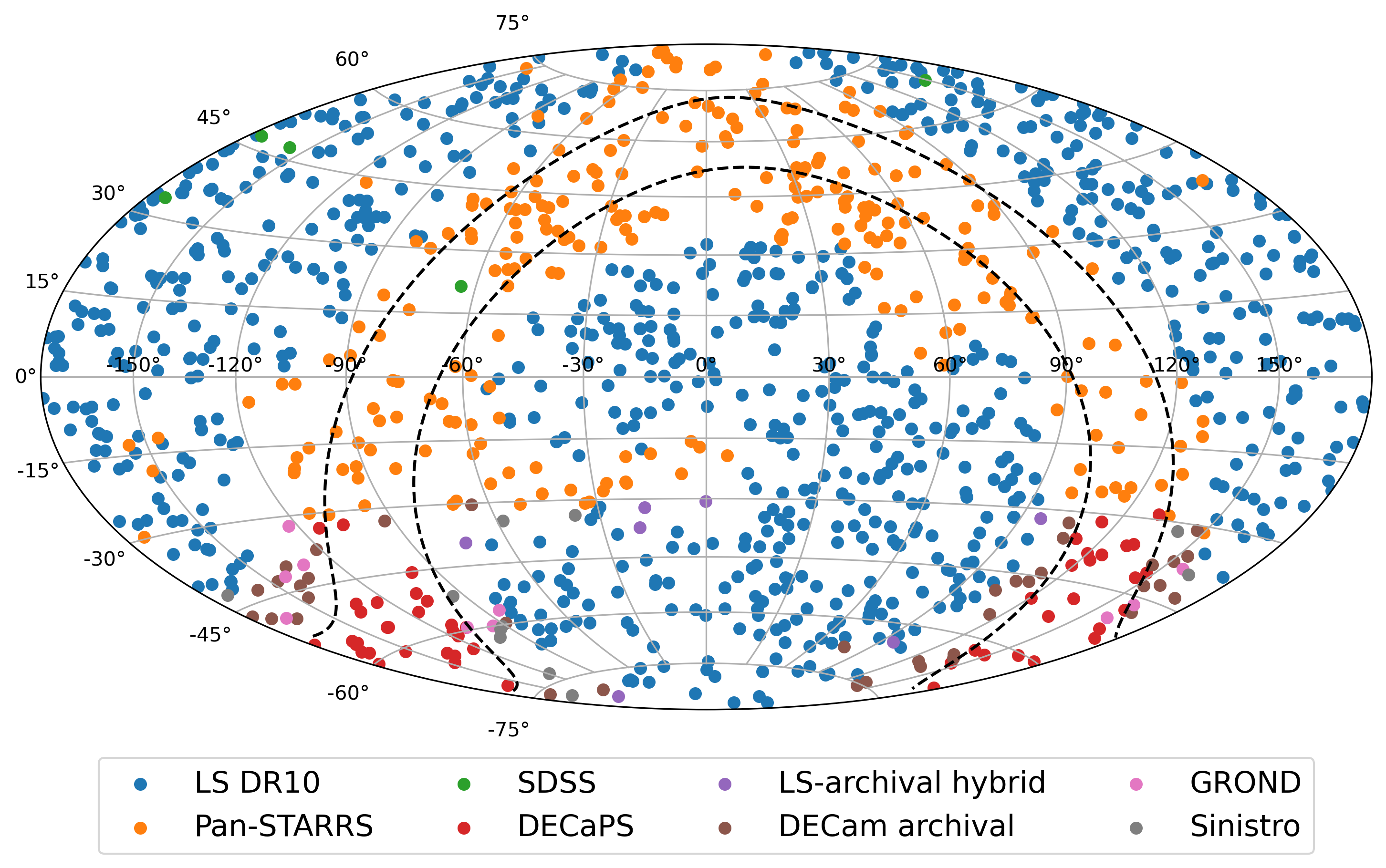}
    \caption{Spatial distribution of BASS AGNs in equatorial coordinates adopting a Hammer projection, colored by the source of the corresponding images. Galactic latitudes $b = \pm 10\degree$ are drawn as dashed curves for reference to identify the Galactic plane.}
    \label{fig:spatial distribution}
\end{figure}

A breakdown of the galaxies and the corresponding source of their imaging is shown in Table \ref{tab:sources}, including reference values for the limiting magnitudes and median \textit{r} seeing of the various sources. Their sky locations are shown in Fig.~\ref{fig:spatial distribution}. 

\begin{table}[]

\resizebox{0.5\textwidth}{!}{ 
\setlength{\tabcolsep}{3pt} 
\centering
\begin{threeparttable}

\caption{Sources of BASS AGNs imaging data.}
\label{tab:sources}
\begin{tabular}{l|rr|rrrr|r}
Image source             & \multicolumn{2}{c}{Galaxy } & \multicolumn{4}{c}{Median 5$\sigma$ limit}            & Median \textit{r} \\ 
  & \# & (\%) & \textit{g} & \textit{r} & \textit{i} & \textit{z} & seeing ('') \\ \hline\hline
LS DR9\tnote{a}  & 200 & (16.8)  & $23.9$ & $23.5$ & -- & $22.5$ & 1.2  \\
LS DR10\tnote{b}  & 607 & (51.0)  & $24.5$ & $24.0$ & $23.7$ & $23.0$ & 1.2  \\
Pan-STARRS DR2           & 261 & (22.0)  & $23.3$ & $23.2$ & $23.1$ & $22.3$ & 1.2  \\
SDSS DR12                & 5   & ( 0.4)  & $23.1$ & $22.7$ & $22.2$ & $20.7$ & 1.2  \\
DECaPS DR2               & 51  & ( 4.3)  & $23.5$ & $22.6$ & $22.1$ & $21.6$ & 1.3  \\
LS-archival hybrid\tnote{c}       & 7   & ( 0.6)  & $24.5$ & $24.0$ & $23.7$ & $23.0$ & 1.2  \\
DECam archival\tnote{d}          & 37  & ( 3.1)   & $23.5$ & $23.0$ & $22.7$ & $22.0$ & 1.2  \\
GROND\tnote{e}           & 10  & ( 0.8)  & $23.6$ & $23.6$ & $22.9$ & $22.7$ & 1.5   \\
Sinistro\tnote{f}        & 11  & ( 0.9)  & $24.2$ & $23.5$ & $23.1$ & $22.3$ & 2.9 \\ \hline
\textbf{Total}           & \textbf{1189} & \textbf{100}
\end{tabular}
\begin{tablenotes}\footnotesize
\item[a] LS DR9 depths for northern fields (`BASS' and `MzLS').
\item[b] LS DR10 depths for southern fields (`DECaLS').
\item[c] Since two out of the three bands come from LS DR10, we cite the same depth and seeing values for this source.
\item[d] Given the archival nature of these files, we conservatively list depths about 1 mag shallower than LS DR10.
\item[e] 5$\sigma$ limit at 520 seconds of exposure, adapted from Table 2 of \cite{greiner2008_grond}.
\item[f] 5$\sigma$ limit at 1500 seconds of exposure. Median \textit{r} seeing estimated from the files.
\end{tablenotes}
\end{threeparttable}
}

\end{table}

\subsection{GROND observations}\label{sec:data_GROND}

Ten BASS AGN were imaged using the GROND instrument on the MPE 2.2m at La Silla Observatory between August 2021 and January 2022 (CN2020A-77, PI E. Treister). Each galaxy was observed between four and five times in \textit{grz} (six cases) or \textit{gri} (four cases) for combined exposures between $\approx$1500--9600 seconds, depending on the brightness of the source, achieving a similar depth to the other objects (see Table \ref{tab:sources}) and allowing one to probe faint features such as potential tidal tails. The reduction of raw science images and calibrations, along with the final co-addition of the reduced products, was performed using THELI \citep{theli}. 

\subsection{Sinistro observations}\label{sec:data_Sinistro}

Eleven galaxies were observed using the Sinistro camera on 1-m telescopes as part of the Las Cumbres Observatory Global Telescope Network (LCOGTN) during the second semester of 2023 (CLN2023B-006, PI F. Bauer), generally with individual objects and images taken on different nights. For each of the galaxies, we took three exposures for a combined total of 1500 seconds in each of \textit{grz} bands, allowing us to achieve a similar image depth as compared to the other objects already in our dataset. We adopted a 10$"$ dither between exposures to account for bad pixels when stacking. We downloaded reduced images from the LCOGTN science archive, which were processed from the raw exposures using the best/most-current calibration files (bias/dark/flat) by the LCOGTN BANZAI pipeline \citep{banzai}. Strong, large-scale fringing patterns were found in all \textit{z} exposures, regardless of the telescope and night. No dedicated fringing calibration observations were taken or made available to correct for this, leaving as our only option to partially remove it using the dithered images and threshold the elevated background out during the image generation process (see Appendix~\ref{sec:Sinistro_fringe})

\subsection{Image generation}\label{sec:data_image_generation}

We constructed composite RGB color images following the methodology devised by \citet{Lupton2004}, whereby the \textit{z}-band (or \textit{i}-band in the event of missing \textit{z}-band) was mapped to the red channel, \textit{r}-band to the green channel, and \textit{g}-band to the blue channel. We adopted a color balance comparable to the one adopted by the SDSS image cutout service, which allowed for features such as Voorwerpjes (VPs) \citep[e.g.,][]{lintott2009_vp} and dust lanes to appear clearer in the imaging.

\begin{figure}[]
        \subfloat[BAT 280 - Shallow stretch]{
            \includegraphics[width=.48\linewidth]{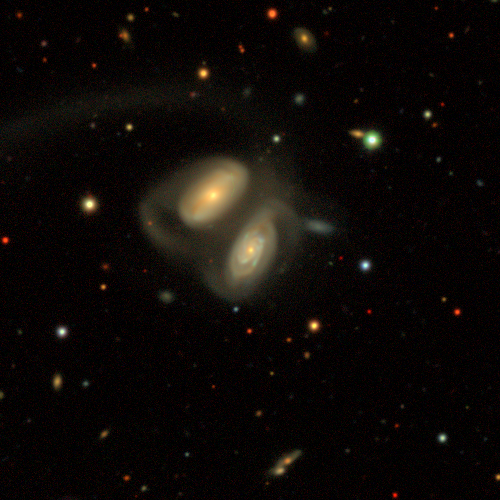}%
        }\hfill
        \subfloat[BAT 280 - Deep stretch]{
            \includegraphics[width=.48\linewidth]{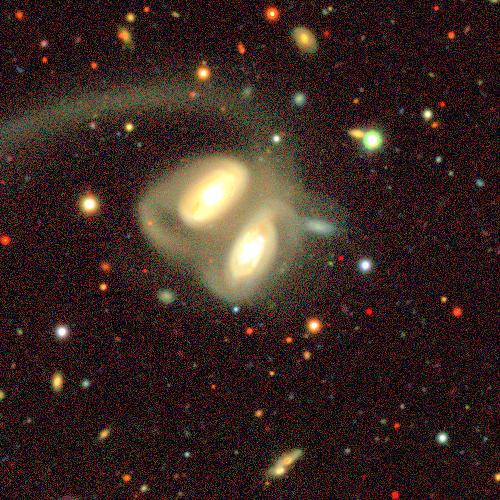}%
        }\\
        \subfloat[BAT 717 - Shallow stretch]{
            \includegraphics[width=.48\linewidth]{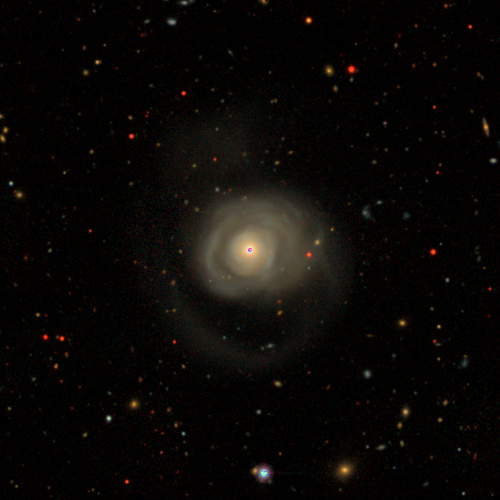}%
        }\hfill
        \subfloat[BAT 717 - Deep stretch]{
            \includegraphics[width=.48\linewidth]{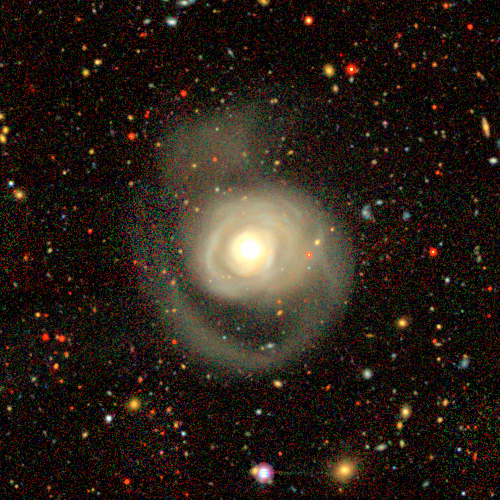}%
        }
        \caption{Shallow and deep stretch comparison highlighting brighter and fainter features, respectively, for BAT 280 (top) and BAT 717 (bottom).}
        \label{fig: stretches}
    \end{figure}

As a means to provide volunteers with more information when examining the subjects, we generated two color images for each galaxy using distinct arcsinh stretches; a shallow stretch to probe details in the nucleus and other bright features and a deep stretch to highlight the surroundings and fainter features such as tidal structures\footnote{Note that the shallow stretch is comparable to what SDSS produces; see, e.g., \newline \url{http://skyserver.sdss.org/dr17/VisualTools/navi}}. This combination aided in better disentangling certain classifications, e.g., regarding galaxies with disturbed morphologies, while maintaining the integrity of central features in the shallow-stretch image. Figure \ref{fig: stretches} shows a comparison between images with different stretches, for which the tidal tails become very apparent in the deep-stretch image.

For the 108 sources that lie in $|b|{<}10^{\circ}$ and have $A_{\rm V}{>}1$, primarily objects imaged with DECaPS and PanSTARRS, we dereddened the images to better represent the intrinsic colors of the galaxy unaffected by Milky Way extinction, for which we used the extinction laws of \cite{CCM89} and the MW dust maps from \cite{planck_dust}. A few examples of this correction are shown in the Appendix~\ref{sec:deredden}. 

In making hybrid images using two bands from LSDR10 and one archival band from DECam, due to uncertainties in zeropoints, we manually scaled the archival band to generate a color image that is generally indistinguishable from the nominal three-band LSDR10 images.

To mitigate the fringing patterns in the red channel of images made from our Sinistro observations, we set stricter floors for clipping the \textit{z}-band images to limit the faint artificial emission in that channel. Although this clipping performed well in removing the fringes, it came at the cost of turning real fainter features somewhat less red. Regarding the morphological assessment, the fringing pattern is generally only a distracting nuisance, primarily seen in the deep-stretch images. We removed the patterns to avoid confusion among classifiers, and stress that this modification was limited to only 11 objects.

\section{Zooniverse project}\label{sec:zooniverse}

One of our objectives is to compare morphological classifications to the broader, commensurate galaxy population, as sampled by the  Galaxy Zoo DECaLS (GZD project \citep{Walmsley2022_gzd}. The GZD-5 decision tree already allows for a relatively detailed and specialized insight into a galaxy's morphological properties, which we implemented with only a few slight modifications based on initial testing using the Zooniverse Project Builder (ZPB)\footnote{Our project is hosted in the following link: \newline \url{https://www.zooniverse.org/projects/maiguelp/the-many-shapes-of-bass}}. We implemented these small deviations from the original GZD-5 workflow to better suit our sample and scientific goals, with the changes including:

\begin{itemize}
    \item In the first task, splitting 
    the ``Star or artifact'' option into 
    ``Point-like'' and ``Artifact/Image problem''. 
    \item Adding a task that asks about faint extended emission and its shape for point-like sources.
    \item Requiring an evaluation of bulge size for every disk galaxy and adding a ``Can't tell'' option. This supplanted and extended the original task related to bulges in edge-on galaxies, which was removed.
    \item Adding ``Clumpy'', ``Possible Voorwerp'', ``Saturated/Blue nucleus'', ``Nearby companion'' and ``Images can be improved'' options to the final question, which asks about extra features and is multiple selection.
    \item Adding a comment section at the end of the decision tree to include additional relevant information. 
\end{itemize}

\begin{figure*}
    \centering
    \includegraphics[width = 0.8\linewidth]{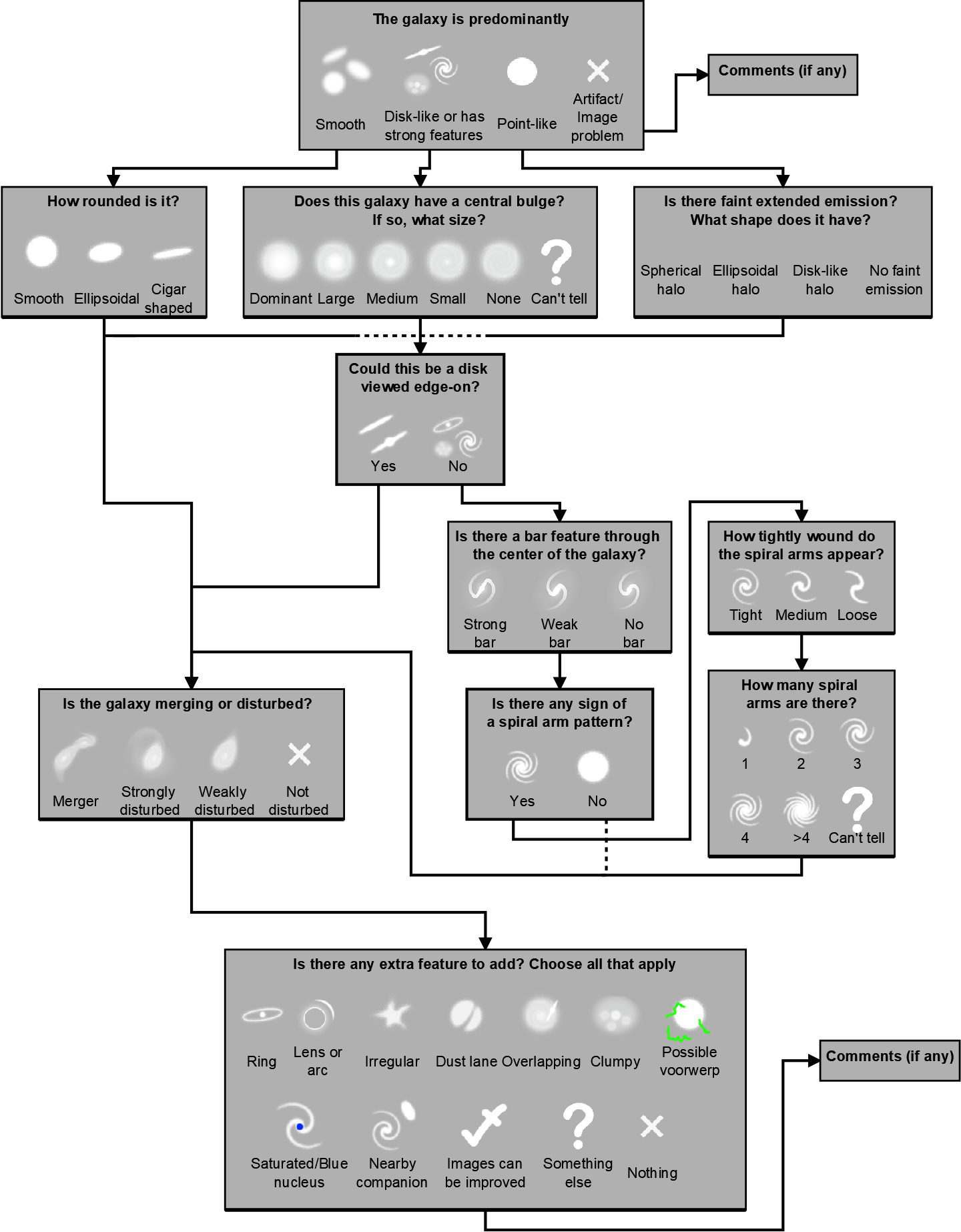}
    \caption{Decision tree implemented in our Zooniverse project.}
    \label{fig:tree}
\end{figure*}

To aid the volunteers in their classifications, each question (or task, in Zooniverse jargon) had its own help section with thorough explanations, pointers, and various examples to guide the tasks. The decision tree and the icons accompanying the answers can be seen in Fig.~\ref{fig:tree}. Most of the aforementioned icons were taken from the GZD-5 project, while others were adapted from pre-existing icons. 

\begin{figure}
    \centering
    \includegraphics[width = 0.95\linewidth]{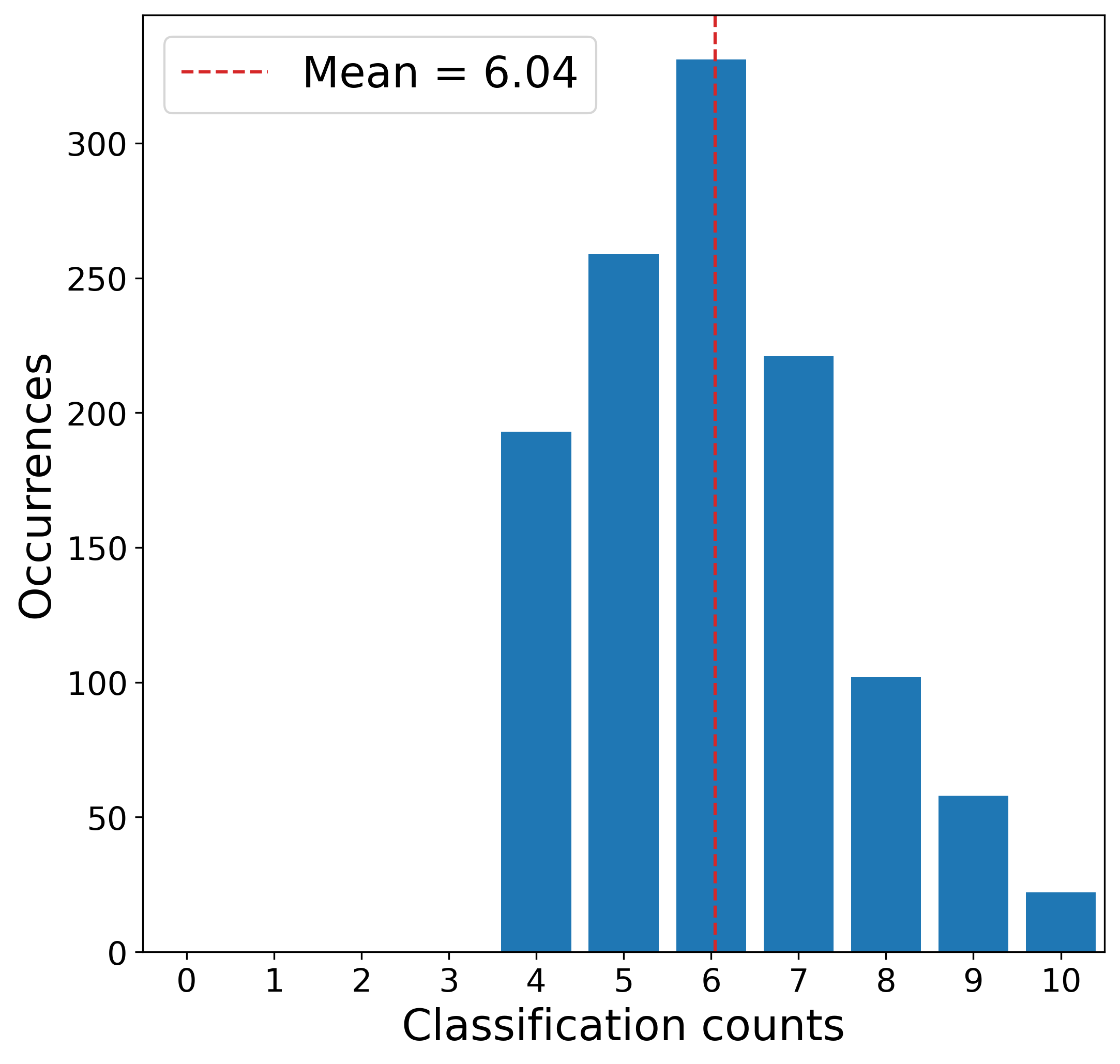}
    \caption{Classification count distribution for the objects uploaded in our project as of January 2025.}
    \label{fig: count distribution}
\end{figure}

Our goal was to reach 5--10 volunteer classifications per object, as this threshold was considered an approximate minimum to allow for meaningful consensus. We only partially achieved this, and thus present our results with the current classifications. \footnote{It should be noted that the project operated semipublicly (`word of mouth'), whereby it was directly shared with interested parties, via private communications or during conference talks, who were, in turn, encouraged to do the same.} In total, there were $>$150 distinct volunteers, although the top seven, all trained scientists, contributed with $\approx$90\% of the classifications. The more prolific volunteers became collaborators of this paper, having made significant contributions to the classification database. The distribution of classification counts can be seen in Fig.~\ref{fig: count distribution}, showing that all galaxies have been classified at least four times, with an average of about six classifications per galaxy.

\section{Comparison with Galaxy Zoo DECaLS} \label{sec:gzd}

The GZD project performed a morphological analysis on $>$300,000 unique objects using images taken by DECam. We use this sample to investigate whether there is any particular morphological preference among galaxies which host AGN. The intentionally similar decision tree (``taxonomy'') and sheer size of GZD make it a valuable sample to reference against BASS. To prevent the control sample from including AGN, we identify and remove them by cross-matching GZD with the MILLIQUAS catalog \citep{Flesch2023_milliquas} (which includes quasars identified in multiple wavelengths and across decades of literature) and with the BASS catalog itself, which resulted in the exclusion of 11037 galaxies or around 3.5\% of the sample.

\subsection{Merging the GZD samples}\label{sec:gzd_merge}

The classification of 313,789 unique galaxies in the GZD sample were split among two phases of GZD, namely GZD-1-2 and GZD-5. GZD-1-2 includes an average of 39 classifications for $>$90k galaxies, while GZD-5 averages 15 classifications across $\approx$250k galaxies; there is an overlap of $\approx$30k objects. The split in versions arises due to small modifications to the decision tree \citep{Walmsley2022_gzd}, which was changed slightly to be more sensitive to bars and mergers for GZD-5 due to the more refined classification. Comparisons between BASS and GZD-5 will be the most straightforward regarding the decision tree, since we based our tree on the GZD-5 template. However, incorporating the classifications from GZD-1-2 raises the average classification count to $\approx$23 while adding 60k new galaxies to the pool, provided that we align the project workflow differences.

Given the scope of our study, the most problematic task is comparing results related to a merger classification. The question itself in GZD-1-2 is worded around merging and tidal debris, and gives the options ``Merging'', ``Tidal debris'', ``Both'' and ``Neither''; on the other hand, GZD-5 used the terminology of merging and disturbances, with options ``Merging'', ``Strongly disturbed'', ``Weakly disturbed'' and ``None''. It is relatively straightforward to map the GZD-1-2 votes for ``Merging'' and ``Both'' to GZD-5's ``Merging'', and ``Neither'' to ``None''. However, to assign the ``Tidal debris'' votes from GZD-1-2 across the degrees of disturbance in GZD-5, we took inspiration from the relation mentioned by \cite{Masters2012_weak_bars} regarding bars, which has been extensively used since then. Although GZD-1-2 only had ``Yes'' and ``No'' options when asking about bars, it was noticed that galaxies where the vote fraction $p_{bar}$ complied with $0.2 < p_{bar} \leq 0.5$ corresponded to weak bars when contrasted with expert analysis; conversely, $p_{bar} > 0.5$ could be interpreted as strong bars and $p_{bar} \leq 0.2$ as lack of bars altogether. We implemented a similar approach for tidal debris. When $p_{tidal} > 0.5$, its votes were counted toward ``Strongly disturbed'', while when $0.2 < p_{tidal} \leq 0.5$, the votes went to ``Weakly disturbed'', and if $p_{tidal} \leq 0.2$, they went to ``Not disturbed''. We confirmed the viability of this conversion after performing a visual inspection on a subsample of GZD images.

\subsection{Establishing a comparison sample}\label{sec:gzd_comparison_sample}

Before comparing BASS and GZD galaxies, we want to ensure that any strong variations introduced by distinct selection effects among the two samples are mitigated. At a minimum, we would like to control for host galaxy stellar mass and distance. Ideally, we could also consider star formation rate and history, environment, and gas content among other quantities, although such a detailed level of control is beyond the scope of this study. As a proxy for stellar mass, we adopt the $i$-band (absolute) magnitude, which measures the galaxy's redder stellar content that is less affected by recent star formation and line-of-sight extinction.

\begin{figure}
    \centering
    \includegraphics[width = \linewidth]{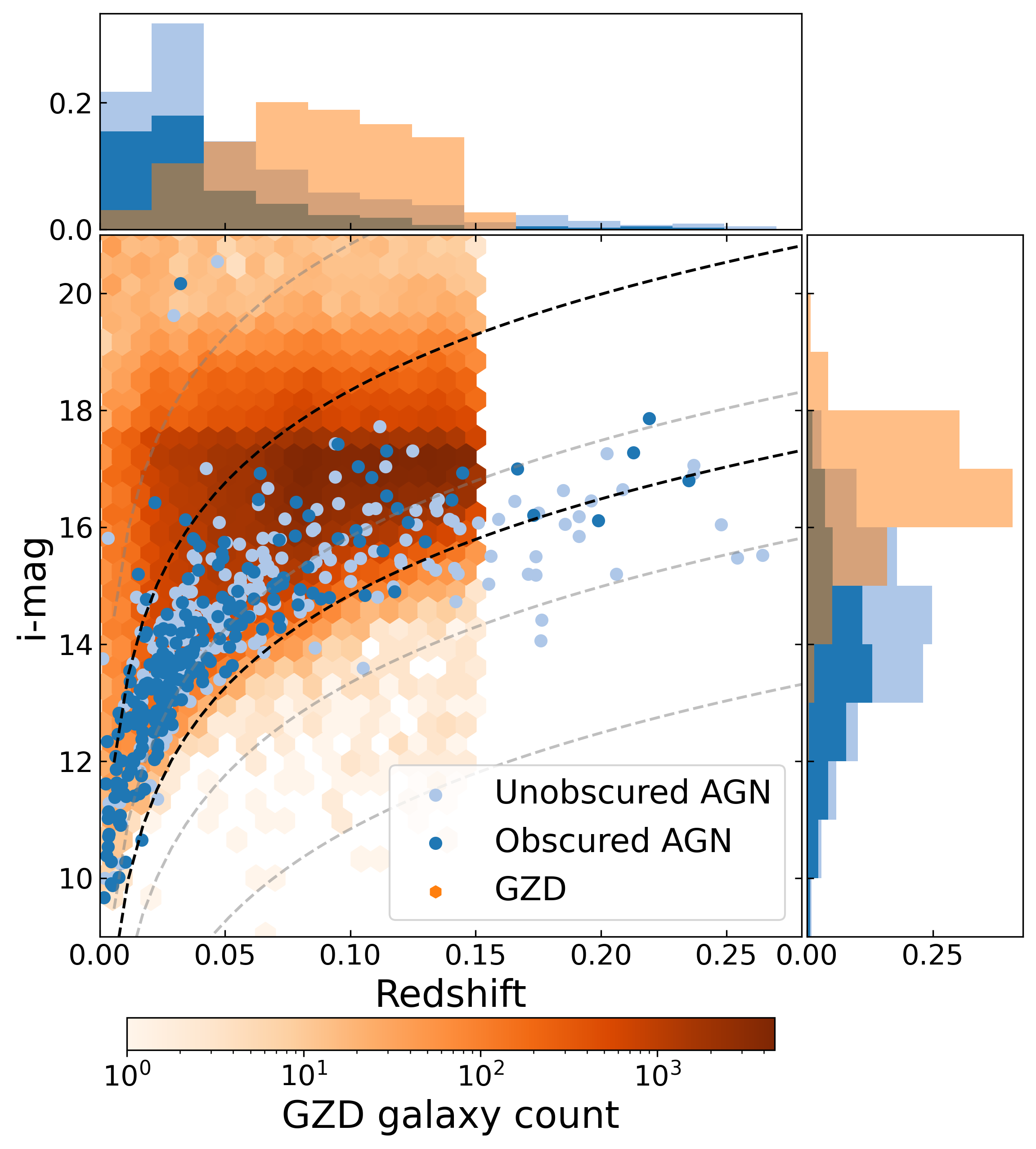}
    \caption{Observed redshift and \textit{i}-mag distributions for BASS Seyferts and GZD galaxies. Only galaxies with \textit{i}-mag measurements are included. Note that although the BASS Seyferts in general cover a redshift range up to $z = 0.52$, for visualization purposes we limit the plot to $z < 0.28$, where $\sim98\%$ of BASS Seyferts are accounted for. The dashed curves represent constant absolute magnitudes, with most of the BASS sample falling between -20 and -23.5 as represented by the black curves.}
    \label{fig: pre-match}
\end{figure}

Considering the full samples, BASS and GZD have significantly different redshift and magnitude distributions, as shown by Fig.~\ref{fig: pre-match}, which could strongly skew the classification counts toward certain morphologies if not considered. The match in redshift is straightforward, but the match in (absolute) magnitude requires us to address the contribution of AGN to the total flux. In particular, with matched redshifts, the BASS sample exhibits brighter magnitudes compared to the GZD sample, as a result of the AGN dominating the measured flux and is thus not representative of the host galaxy's contribution, which would otherwise be tantamount to a tracer of stellar mass and star formation rate. For this reason, and only to arrive at a matched comparison sample, we limit the matching of the BASS sample only to obscured AGN where their contribution to the measured magnitude is considered insignificant. By matching GZD to this subset of AGN and then comparing to the entire AGN sample regardless of obscuration, we make the implicit assumption here that obscured and unobscured AGN have similar host property distributions, specifically the stellar mass and star formation rate (SFR). Type 1 and 2 AGN have been found to exhibit similar SFRs \citep[e.g.,][]{Zou2019_mass_type1_2, Mountrichas2021_mass_type1_2}, and although the same studies also report type 1 AGN being slightly less massive than their type 2 counterparts, said differences in stellar mass range from statistically insignificant ($\sim1\sigma$) to quantitatively small either way ($\sim$ 0.2--0.3 dex). We validate this assumption in Appendix \ref{sec:results_BASS_NH}. Here we separate between obscured and unobscured AGN based on optical Seyfert classification, where obscured AGN correspond to sources classified as Sy1.8, Sy1.9 or Sy2, and unobscured AGN are associated with Sy1, Sy1.2 and Sy1.5; we also show in Appendix \ref{sec:results_BASS_NH} that our conclusions remain valid if we separate obscured AGN based on X-ray derived hydrogen column densities, but note that this sample is more limited due to the availability of $N_\mathrm{H}$ measurements. Considering this, we matched the \textit{i}-mag distributions using photometry from \cite{Koss2011_BASSMorphs} for the BASS and from SDSS for GZD.

To achieve a matched comparison sample, we select objects randomly from the GZD sample, applying the distribution of the obscured BASS AGN as a prior. For this comparison, we also capped the BASS sample at $z{=}0.15$, to align with the strict cutoff from GZD. There are multiple possible draws to assemble a matched GZD subsample. To account for uncertainties related to this, we generated 6000 matched GZD subsamples, computed their morphological class fractions, and took the median of each category, while the 1-$\sigma$ errors were assessed from the respective distributions.

\begin{figure}
    \centering
    \includegraphics[width = \linewidth]{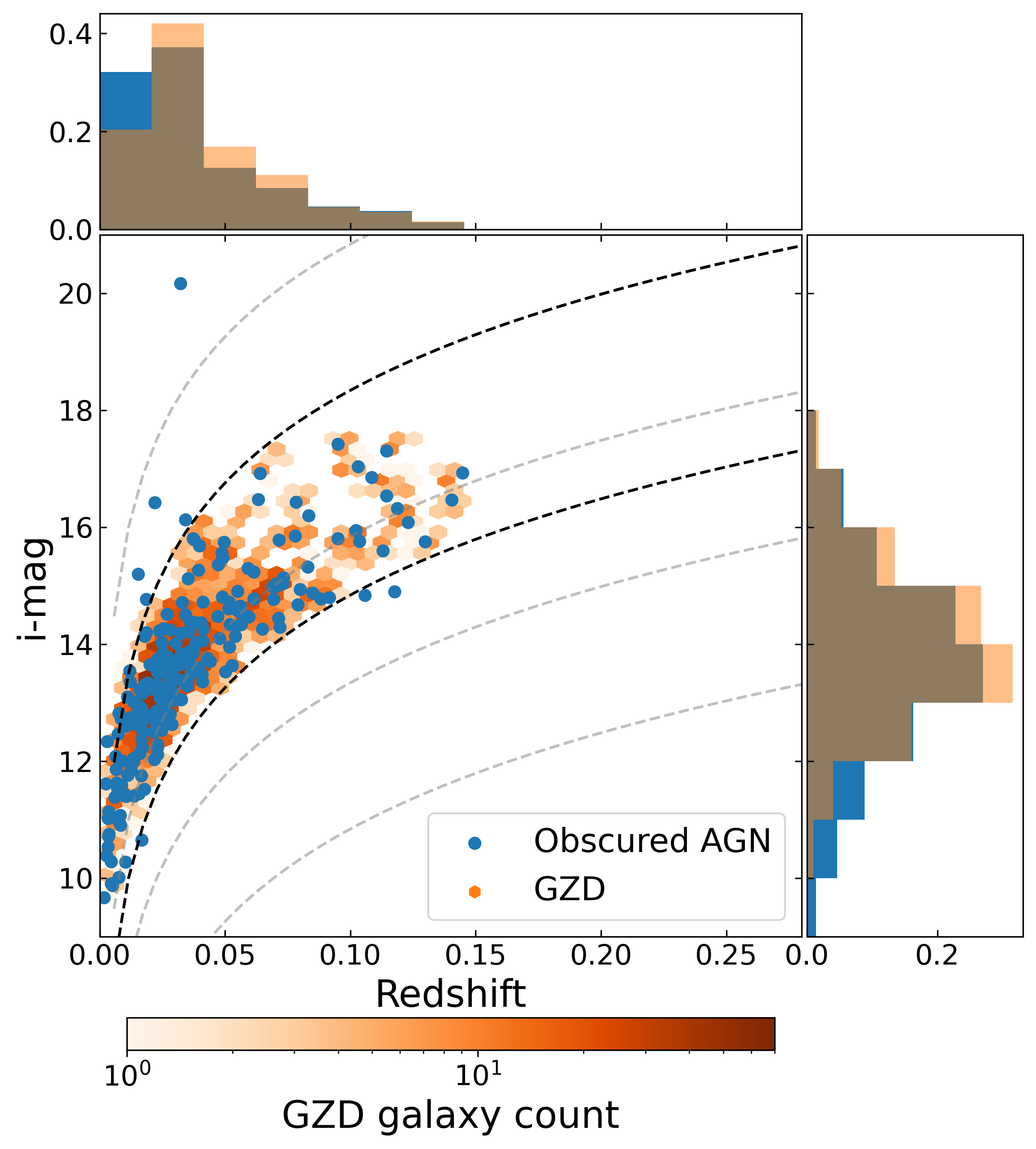}
    \caption{Redshift and \textit{i}-mag distributions for the obscured BASS and one of the matched GZD subsamples. Only galaxies with \textit{i}-mag measurements are included. Note that although the BASS Seyferts, in general, cover a redshift range up to $z = 0.52$, for visualization purposes, we limit the plot to $z < 0.28$, where $\sim98\%$ of BASS Seyferts are accounted for. The dashed curves represent constant absolute magnitudes, with most of the BASS sample falling between -20 and -23.5 as represented by the black curves}
    \label{fig: post-match}
\end{figure}

The final GZD comparison subsamples consist of $\sim$2550 galaxies, matched to the redshift and magnitude distributions from 215 obscured AGN host galaxies from BASS. To summarize, BASS galaxies needed available \textit{i}-band magnitude measurements, a spectroscopic classification as Seyfert 1.8, 1.9 or 2, and a redshift below $z = 0.15$ to ensure that the AGN itself is obscured and within the GZD redshift intervals. Likewise, GZD galaxies needed available \textit{i}-magnitude measurements and redshift to be eligible for the matching distributions, and to specifically populate the same range of absolute \textit{i}-magnitudes as BASS ($-20$ to $-23$). Finally, we required that GZD galaxies have at least 25 classifications to reduce uncertainties that could arise from a low voting count. Figure \ref{fig: post-match} shows the distributions of the comparable samples in $z$ and \textit{i}-mag. Despite the number of galaxies available in GZD, the overlapping parameter space is not exact (BASS has a higher fraction of brighter and closer galaxies than can be found in GZD), and could be one source of bias at the $\sim$10\% level (based on the histograms in Fig.~\ref{fig: post-match}). 

\section{Classifications}\label{sec:classifications}

We translated the matrix of task responses from Zooniverse into morphological classes based on quorums on a compilation of specific tasks for both BASS and GZD. As a starting point, we adopted the suggested classification fractions from \citet{Walmsley2022_gzd}, that requires a quorum of 60\% in a particular task to indicate a confident consensus around a feature's presence (or absence). We did not require the same threshold of classifications for BASS, since it had a much lower average per galaxy at the time of publication. Somewhat related to lower average vote totals, we identified some subsets of galaxies where secondary traits such as their merger status or disk orientation ended up being a source of disagreement; for these cases, in both BASS and GZD, we chose to designate them to a clear class as follows conservatively: galaxies with no merging consensus (0.4 $\leq$ \texttt{Merging} + \texttt{Strongly disturbed} $\leq$ 0.6) were treated as having no disturbance (thus we modify \texttt{Merging} + \texttt{Strongly disturbed} < 0.4 to \texttt{Merging} + \texttt{Strongly disturbed} $\leq$ 0.6), whereas no disk orientation consensus (0.4 $\leq$ \texttt{Face-on} $\leq$ 0.6) were treated as being face-on (thus we modify \texttt{Face-on} > 0.4). 

\begin{figure*}
    \centering
    \includegraphics[width= 0.8\linewidth]{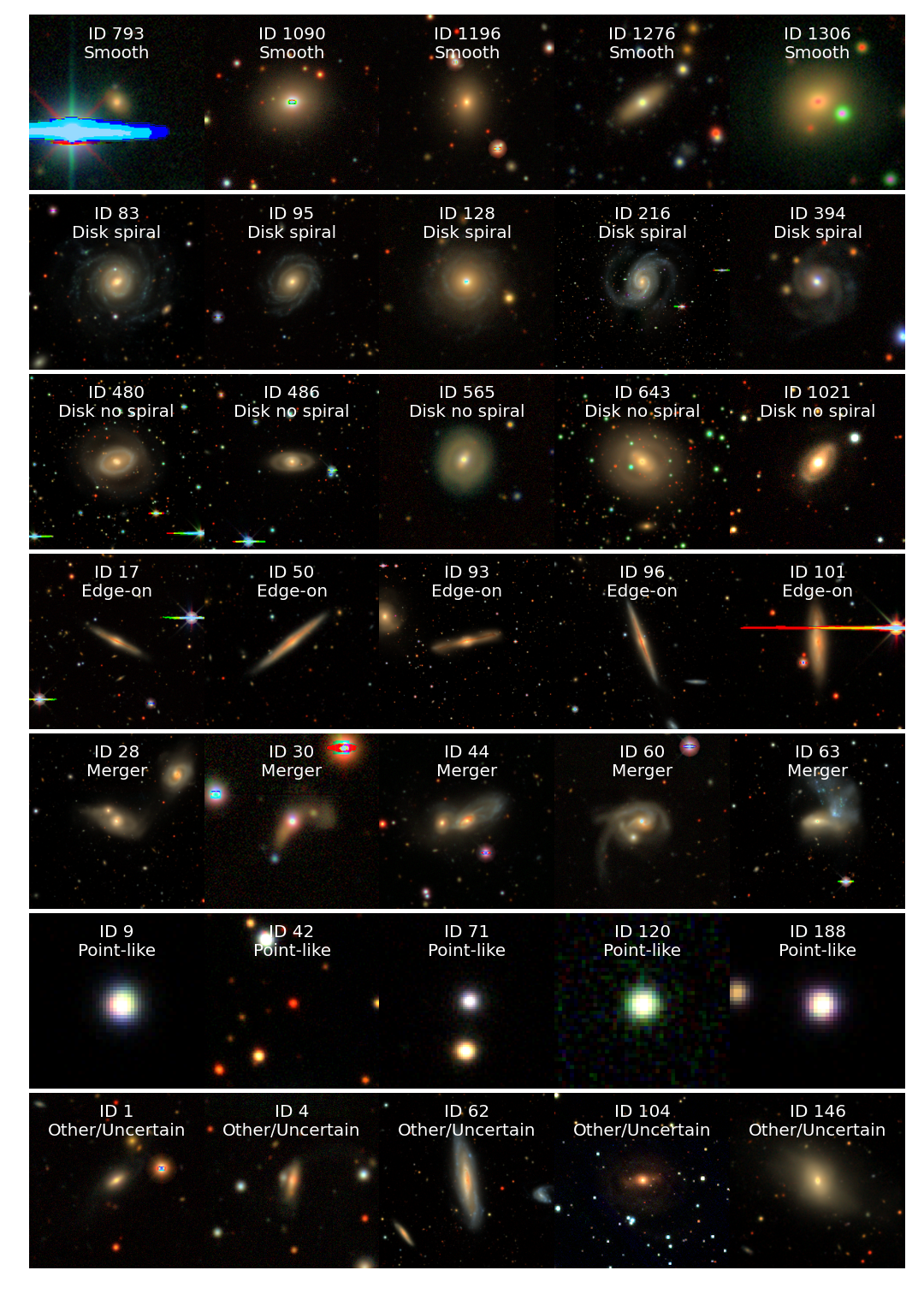}
    \caption{Selection of representative BASS AGN host galaxies for each broad morphological class defined in our classification scheme. From top to bottom: Smooth, disk spiral, disk no spiral, edge-on, merger, point-like, and other-uncertain.}
    \label{fig:compilation}
\end{figure*}

Each class, along with its required features and cuts, is depicted in Table~\ref{tab: criteria}, where a galaxy must satisfy each condition to be considered a member of a certain class. A critical distinction between our workflow and GZD is how we treat point source morphologies. Because our sample includes many powerful AGN, we implement separate classes for point-like galaxies and artifacts in our taxonomy. By contrast, the GZD workflow groups point sources and artifacts together, leaving no way to single out point-like galaxies.
The point-like/artifact class accounts for 0.7\% of all GZD galaxies; based on visual inspection for a random subset, we found that most appear to be true artifacts, thus excluding this entire subset from further analysis. 
At the same time, point-like galaxies comprise 13.5\% of the full BASS sample, with none being artifacts (essentially by design, given the {\it Swift}-BAT pre-selection). Notably, once redshift and magnitude constraints are imposed on the population to match the GZD control sample, only 2.2\% (five sources) in the BASS comparison sample remain as point-like. With this in mind, we stress that their inclusion or exclusion as a category in its own right is not critical to the comparison study. 
Figure~\ref{fig:compilation} presents a selection of representative galaxies belonging to each main class.

\begin{table}
\fontsize{9pt}{9pt}\selectfont
\centering
\caption{Classification criteria for the classes.}
\label{tab: criteria}
\begin{tabular}{@{}ll@{}}
\toprule
Class                     & Criteria                                             \\ \midrule
Smooth                    & \texttt{Smooth} $>$ 0.6                                \\
                          & \texttt{Merging} + \texttt{Strongly disturbed} $\leq$ 0.6 \\ \midrule
Disk-spiral               & \texttt{Disk-like/featured} $>$ 0.6                              \\
                          & \texttt{Face-on} $>$ 0.4                               \\
                          & \texttt{Spiral arms} $>$ 0.6                           \\
                          & \texttt{Merging} + \texttt{Strongly disturbed} $\leq$ 0.6 \\ \midrule
Disk-no spiral          & \texttt{Disk-like/featured} $>$ 0.6                              \\
\multicolumn{1}{l}{}      & \texttt{Face-on} $>$ 0.4                               \\
\multicolumn{1}{l}{}      & \texttt{No spiral arms} $>$ 0.6                        \\
\multicolumn{1}{l}{}      & \texttt{Merging} + \texttt{Strongly disturbed} $\leq$ 0.6 \\ \midrule
Edge-on                   & \texttt{Disk-like/featured} $>$ 0.6                              \\
                          & \texttt{Edge-on} $>$ 0.6                               \\
                          & \texttt{Merging} + \texttt{Strongly disturbed} $\leq$ 0.6 \\ \midrule
Merger-strongly disturbed & \texttt{Merging} + \texttt{Strongly disturbed} $>$ 0.6 \\ \midrule
Point-like                & \texttt{Point-like} $>$ 0.6                            \\ \midrule
Unclassifiable            & \texttt{Artifact} $>$ 0.6                              \\ \midrule
Other-uncertain           & Fails the criteria above                 \\ \bottomrule
\end{tabular}
\end{table}

\section{Reclassifying the other-uncertain}\label{sec:reclassifications}

A non-negligible fraction ($\sim$15--20\%) of the galaxies in the BASS and GZD samples do not fall into the first 6--7 strictly defined classes of Table~\ref{tab: criteria} due to discrepancies or uncertainties among the volunteers. These galaxies are assigned to the `other-uncertain' class, a catch-all category for galaxies that failed to reach sufficient consensus in one or more of the listed criteria. However, one must be careful when interpreting this category because galaxies can arrive at this class due to a lack of consensus on any individual (or more) decision tree question listed in Fig.~\ref{fig:tree}.

To help disentangle these possibilities, to first order, we can consider relaxing one of the criteria to understand what the closest morphological class, and hence what the most likely origin of the uncertainty, might be. For example, a galaxy might not have achieved enough consensus in a specific criterion, i.e., its average merging status was 0.5, but would otherwise fulfill all the other qualifications to be successfully classified as, e.g., a disk Spiral. On the other hand, the object may lack consensus across multiple decision tree nodes (typically stemming from disagreements at or close to the root node).

Considering the above, we found that a plurality of the other-uncertain galaxies in BASS (65 out of 145) failed to achieve sufficient consensus on the first question regarding their broad properties  (\texttt{Smooth}, \texttt{Disk-like/featured}, \texttt{Point-like}, \texttt{Artifact}), which we considered to be a fundamental uncertainty, potentially related to the faintness, size, redshift and/or morphology of the target (e.g., see Sect. \ref{sec:results_BASS}).

Regarding `other-uncertain' BASS galaxies with point-like classifications, we made use of the child-node question about the appearance of any faint extended emission to disambiguate and reclassify galaxies as smooth or disk-like galaxies, or let them remain classified as point-like. The reclassification criteria for point-like galaxies is summarized in Table \ref{tab: repoint}, where lack of faint extended emission (or no consensus on the geometry of said emission) is treated as a generic point-like. Only two galaxies go from point-like to disk under this reclassification, so assigning them to a specific subcategory of disk should not alter the statistics meaningfully. A section of the morphological classifications that we release with this article is shown in Table \ref{tab: morpho_cat}.

\begin{table}
\fontsize{9pt}{9pt}\selectfont
\centering

\caption{Reclassification criteria for point-like galaxies.}
\label{tab: repoint}
\begin{tabular}{@{}ll@{}}
\toprule
New class                 & Criteria                                             \\ \midrule
Smooth                    & \texttt{Smooth} $>$ 0.6                                \\
                          & \texttt{Spherical halo} + \texttt{Ellipsoidal halo} $>$ 0.6 \\ \midrule
Disk                      & \texttt{Disk-like/featured} $>$ 0.6                   \\           
                          &  \texttt{Disk-like halo} $>$ 0.6                        \\ \midrule
Point-like                & Fails the criteria above                               \\ \midrule
\end{tabular}
\end{table}

\begin{table*}[]

\centering
\begin{threeparttable}

\caption{Morphological classifications.}
\label{tab: morpho_cat}
\begin{tabular}{llllll}
BAT ID  & \textit{Swift} name & Counterpart name & Main classification      & Reclassification & Bar strength    \\ \hline\hline
36      & SWIFT J0051.9+1724 & Mrk 1148         & Point-like                & Smooth           &                \\
58      & SWIFT J0111.4-3808 & NGC 424          & Other-uncertain           & Disk no spiral   & Unbarred       \\
184     & SWIFT J0333.6-3607 & NGC 1365         & Disk spiral               & Disk spiral      & Weak           \\
573     & SWIFT J1148.7+2941 & MCG +05-28-032   & Edge-on                   & Edge-on          &                \\
1216    & SWIFT J0043.9-5009 & NGC 238          & Disk spiral               & Disk spiral      & Strong         \\
1632    & SWIFT J2355.9+2555 & VV697            & Merger-strongly disturbed & Merger-strongly disturbed &                \\ \hline

\end{tabular}
\begin{tablenotes}\footnotesize
\item \textbf{Notes:} A portion of the morphological catalog is shown here for guidance regarding the published features. Only galaxies classified and/or reclassified as disk spiral or disk no spiral have the strength of their bar listed in the Bar strength column.

\end{tablenotes}
\end{threeparttable}

\end{table*}

Additionally, for galaxies classified as other-uncertain in GZD, we remove \texttt{Artifact} votes as part of the reclassification process. Even though the sample has already been cleaned up from volunteers who disproportionately voted for artifacts \citep{Walmsley2022_gzd}, some contamination remains from artifact votes. We justify the exclusion of artifact votes in the reclassification process on the grounds mentioned above, as well as the extremely low incidence of artifact sources in the first place.

\section{Morphology relations with AGN and host properties}\label{sec:results_BASS}

\begin{figure*}
\includegraphics[width=1\linewidth]{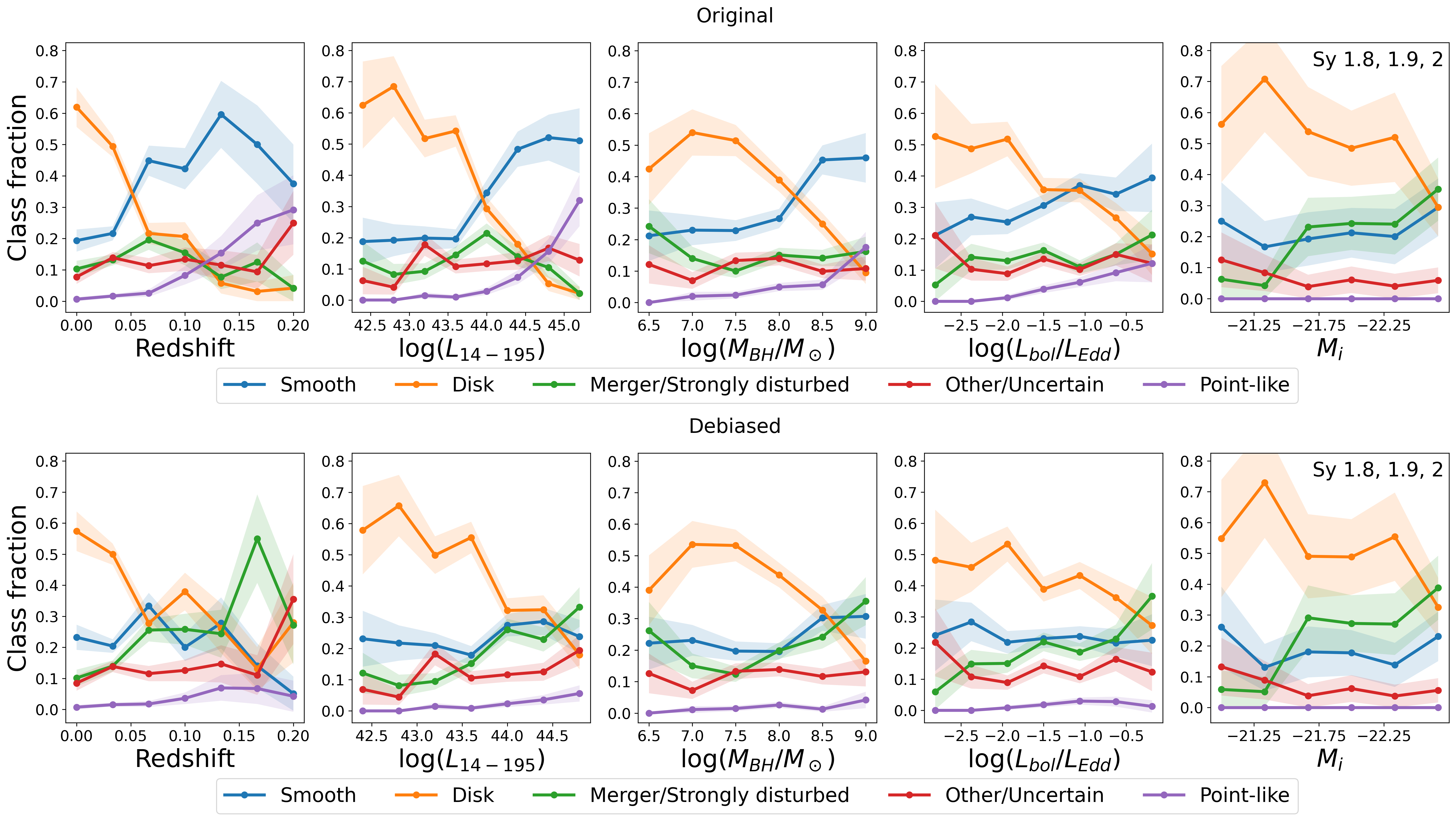}
\caption{Morphological class fraction of BASS AGNs, before (above) and after (below) redshift debiasing, as a function of redshift (left), 14--195\,keV luminosity (middle left), black hole mass (middle), Eddington ratio (middle right), and absolute \textit{i}-magnitude for Sy 1.8-2 types only (right). Shaded regions correspond to 1$\sigma$ confidence. No attempt was made to debias the other-uncertain class. A small percentage of sources lie beyond the limits of the plots, but subsequently have poorly constrained class fractions and were thus cropped out.}
    \label{fig:morph_v_params}
\end{figure*}

To study how the general properties of BASS AGN relate to their morphologies, we examine the wealth of ancillary data available for the sample \citep{Koss2022_BASSDR2_mbh_lbol_agn} and check for trends (well-known or otherwise). To make visualization less cumbersome, we combine the family of disk classes (disk-spiral, disk-no spiral, and edge-on) into a single generic disk class. Additionally, galaxies successfully reclassified with a secondary morphology are counted separately from the other-uncertain class, and are instead considered as members of their secondary morphology class. Finally, we exclude 159 beamed AGN from further analysis to avoid interpretation complications; 92 have point-like morphologies, comprising 69\% of that class in the global BASS sample, and reside at high redshifts. When taking into account available measurements for the individual parameters, this results in a sample of 1015 galaxies for $z$ and $L_{\rm 14-195\,keV}$, 1014 for $L_{\rm bol}$, 960 for $M_{\rm BH}$ and $L_{\rm bol}/L_{\rm Edd}$, and 215 for $M_{\rm i}$ (only Sy 1.8, Sy 1.9 and Sy 2). For details on the measurements of $M_{\rm BH}$ and $L_{\rm bol}/L_{\rm Edd}$, see Section 4.4 of \cite{Koss2022_BASSDR2_mbh_lbol_agn}, although note that we are using values from the DR3 catalog (M. Koss et al. in prep) which include the quantities from \cite{Koss2022_BASSDR2_mbh_lbol_agn} along with new sources.

The morphological dependence with redshift is shown in the top row of Fig.~\ref{fig:morph_v_params}, where the disk galaxy fraction dominates at the lower-end and progressively decreases, while smooth and point-like galaxy fractions dramatically increase their contributions with increasing redshift. These results must first be interpreted in observational terms before astrophysical ones. In particular, when combined with the decreasing apparent angular size of a given galaxy's physical extent with increasing redshift, the roughly fixed spatial resolution of the incorporated imaging surveys will lead to the perceived loss of observed features. Moreover, Malmquist bias, whereby more luminous AGN can be detected out to higher redshifts than weaker ones, will lead to an increasing point source dominance with increasing redshift, in addition to enhancing any luminosity-dependent host galaxy trends. We do not expect the redshift evolution of these features to vary so strongly over a span of only $\lesssim$1--2.5 Gyr \citep[e.g.,][]{ferreira+23}.   

We want to correct the observational trends via redshift-resolution debiasing \citep[e.g.,][]{hart2016_redshift_debias, Walmsley2022_gzd}, but owing to the low numbers of objects and the low numbers of votes for each object in our sample, it is beyond the scope of this work to simulate large numbers of spatially degraded galaxies for reclassification or internally apply the same strategy to the BASS AGN sample as, e.g., \citet{Walmsley2022_gzd} did for the GZD sample. We note, however, that the BASS and GZD samples share a similar parameter space, particularly when it comes to the GZD galaxies used to assess debiasing weights. Hence, we will apply the relative corrections shown in Fig. 10 of \citet{Walmsley2022_gzd} for the smooth, disk, merger-strongly disturbed, and point-like classes, extrapolating these trends from $z{=}0.02$--0.15 to $z{=}0.0$--0.2 (see Appendix~\ref{sec:debias} for details). The debiased trends are shown in the bottom row of Fig.~\ref{fig:morph_v_params}, allowing us to investigate the residual dependencies of morphology on redshift, hard X-ray luminosity, black hole mass, accretion rate (probed by the Eddington ratio) and absolute \textit{i}-band magnitude $M_{\rm i}$. For absolute \textit{i}-magnitude we only examine trends for AGN types 1.8--2, where contamination to the host light is less problematic.

The strongest trend, both in the original and debiased plots, appears to be the decline in the fraction of disk galaxies, which is seen to some extent in every tracer. We argue that this is likely due to Malmquist bias in the primary X-ray luminosity selection, whereby smaller SMBHs and/or lower accretion rates are associated with disk galaxies. This trend is weakest, or at least the most uncertain, in the $M_{\rm i}$, which primarily traces host galaxy mass.

Once debiased, the strong smooth galaxy trends essentially flatten out across all tracers, with a typical fractional contribution of $\sim$20--30\% in $L_{\rm 14-195\,keV}$, $M_{\rm BH}$, and $L_{\rm bol}/L_{\rm Edd}$ and $\sim$15\% in $M_{\rm i}$. The only decline appears at $z>0.15$, where uncertain classification grows. The percentage drop in $M_{\rm i}$ is interesting and may suggest that smooth galaxies are slightly less obscured on average.

Intriguingly, the relative fraction of the debiased merger-strongly disturbed class appears to trend upward across nearly all indicators. This could be a result of interactions and mergers driving increased amounts of gas to the central SMBH, thereby fueling more luminous AGN phases \citep[e.g., ][]{treister2012_merger_lumagn, Ricci2017_Merger_growth}, coupled with Malmquist bias preferentially selecting only higher luminosity (and hence higher $M_{\rm BH}$ and $L_{\rm bol}/L_{\rm Edd}$) objects. Finally, interactions and mergers will be more common in higher-density environments, which are preferentially dominated by more massive (and hence higher $M_{\rm i}$) galaxies, explaining the trend in the debiased $M_{\rm i}$ panel.

Finally, we note that the strong evolution of the point-like category before debiasing is related to changing ratios of resolved to unresolved emission with distance, with the latter including both the AGN and an increasing fraction of host light. The point-like class fraction is strongly suppressed after debiasing, as we might expect given the luminosity range of the AGN; such AGN are generally not bright enough to completely overwhelm the light from their relatively massive hosts. 

\section{Color-magnitude diagrams}\label{sec:results_colormag}

\begin{figure*}
    \centering
    \includegraphics[width=\linewidth]{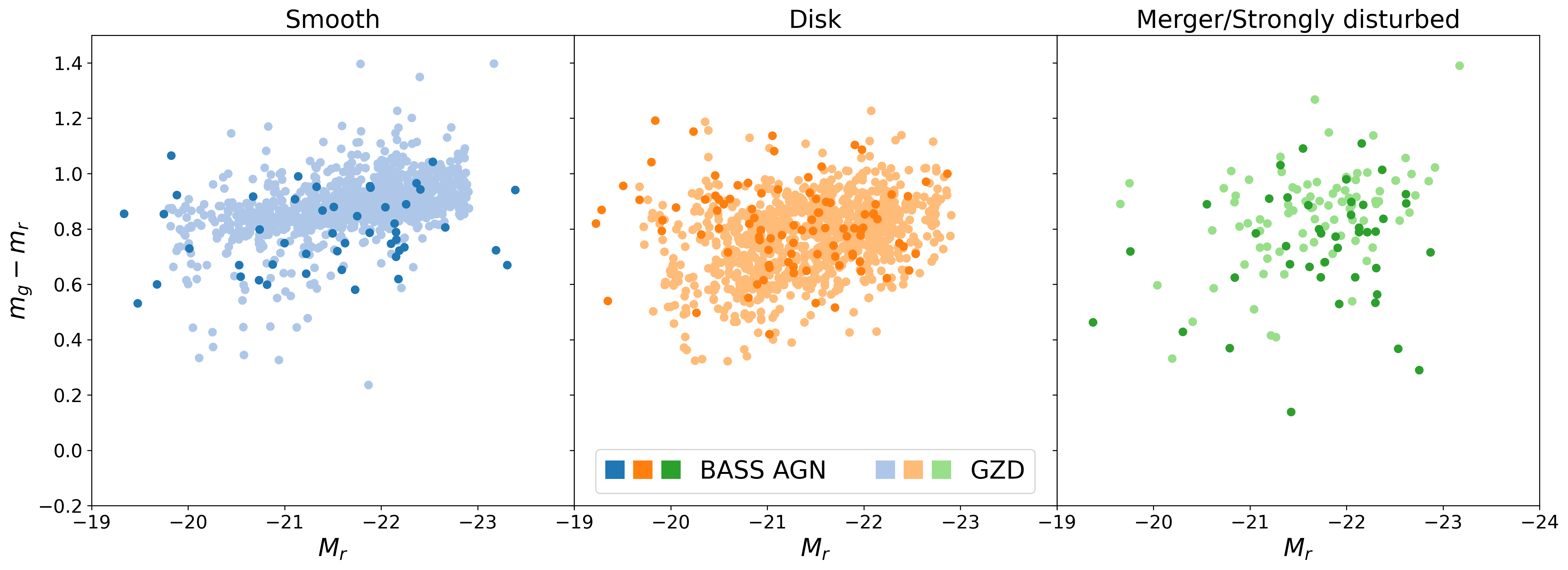}
    \caption{Color-magnitude diagram for BASS AGNs and GZD galaxies, separated by morphological class. Only BAT AGNs classified as Sy1.8, Sy1.9, and Sy2 are included.}
    \label{fig:cm_plot}
\end{figure*}

Color-magnitude diagrams can be insightful tools for examining the parameter space that certain galaxies inhabit, with both properties closely related to visual morphologies. We do not attempt in this work to deblend host galaxy and AGN light, and therefore only investigate this relation for BASS AGN with predominantly narrow emission lines (Sy1.8--2 classes). In Fig.~\ref{fig:cm_plot} we show color-magnitude diagrams for both BASS AGN and a matched sample of inactive galaxies from GZD, with photometry mainly sourced from SDSS for both samples and complemented with measurements from \cite{Koss2011_BASSMorphs} for BASS AGN. 

Among smooth galaxies, the BASS AGN tend to be uniformly distributed over a somewhat larger area in color space than inactive GZD galaxies, with a much higher proportional fraction skewing to bluer colors, either due to the contributions of the AGN itself or a somewhat younger (green valley) stellar population. This trend is not seen for disk galaxies or merger-strongly disturbed systems, with BASS AGN exhibiting similar distributions in color-magnitude space to their GZD inactive counterparts. That said, there appears to be a very mild tendency for AGN hosts in fainter ($M_{\rm r} \gtrsim -21$) disks to exhibit redder colors than their inactive counterparts. Although the trend amounts to perhaps 0.2 in color, several mechanisms could be responsible for the observed redness at the fainter regime, such as increased dust extinction, suppressed star formation, and/or different fueling mechanisms for the AGN.

\section{Comparisons with GZD}\label{sec:results_comp}

% One immediate application of the morphological catalog we assembled for the BAT AGN is to examine morphological tendencies between galaxies hosting AGN and inactive galaxies

\subsection{Common BASS and GZD objects}\label{sec:results_comp_common}

We identified 105 common objects that were classified by both our volunteers and those from GZD, to evaluate the degree of similarity of categorization when dealing with the same galaxies. These common BASS objects are not necessarily part of the comparable BASS sample of 215 galaxies for whom the matching was applied, because no constraint on $N_\mathrm{H}$ or \textit{i}-mag was put in place; likewise, there was no restriction regarding voting count for GZD on this specific comparison.

\begin{figure} 
    \centering
    \includegraphics[width = 0.9 \linewidth]{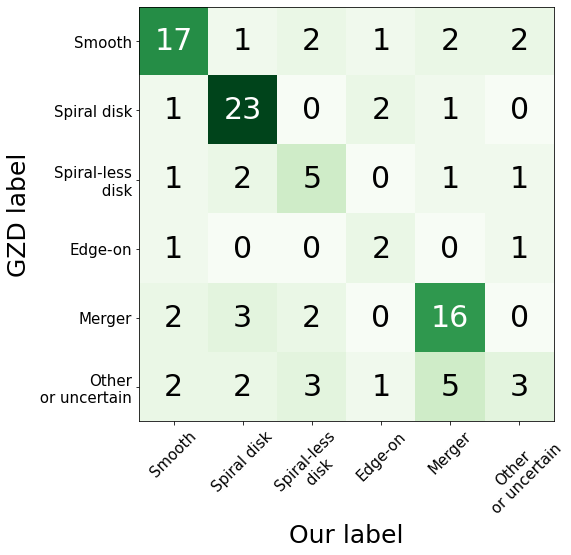}
    \caption{Confusion matrix comparing classifications for common BASS/GZD galaxies.}
    \label{fig:confusion}
\end{figure}

We graphically illustrate the classification consistency using the confusion matrix pictured in Fig.~\ref{fig:confusion}.
If we define successful classifications as ones that are not labeled as other-uncertain, we find that 63 out of 89 successful GZD classifications lie in the diagonal overlap regions, meaning $\sim$70\% of GZD's successful consensus classifications were labeled consistently by both classifiers.
Additionally, we see that our classifications attained higher levels of sufficient consensus (following the criteria in Table \ref{tab: criteria}) than GZD, as evidenced by the lower percentage of other-uncertain galaxies: 6\% in BASS versus 15\% in GZD.

The other-uncertain category here formally includes point-like and unclassifiable objects. However, these amounted to only 2 point-like sources in BASS and 1 artifact\footnote{The latter is BAT 405, also known as UGC 4211. It is very clearly a merger according to the BASS classification and is even a confirmed dual AGN \citep{Koss2023_UGC4211}, but the imaging used in GZD was centered on a star. This probably affected the GZD classification of this source.}  in GZD; thus, their inclusion or exclusion does not meaningfully impact our analysis of this comparison.

The differences in classifications, identified by the off-diagonal elements of Fig.~\ref{fig:confusion}, cannot and should not be explained in terms of ``False Positives'', ``False Negatives'' and the like, because we are not evaluating how a classifier (our volunteers) performed against a gold standard (GZD volunteers) like in the context of a trained machine learning (ML) algorithm. Instead, we compare two separate human classifiers presented with the same physical objects under distinct conditions and with different tools. Namely, the generation of the images followed different approaches, such as the definition of the Field of view (FOV), the availability of two stretches, and choice of coloration. The different amounts of information provided to the respective volunteers can explain most of these disagreements over the same galaxies.

\begin{figure}[]
        \subfloat[BAT 1189 (GZD)]{%
            \includegraphics[width=.32\linewidth]{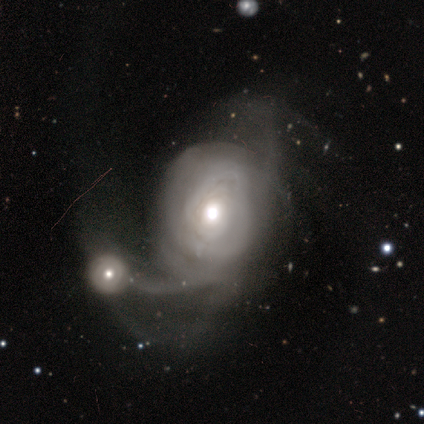}%
        }\hfill
        \subfloat[BAT 1189 (Ours, shallow stretch)]{%
            \includegraphics[width=.32\linewidth]{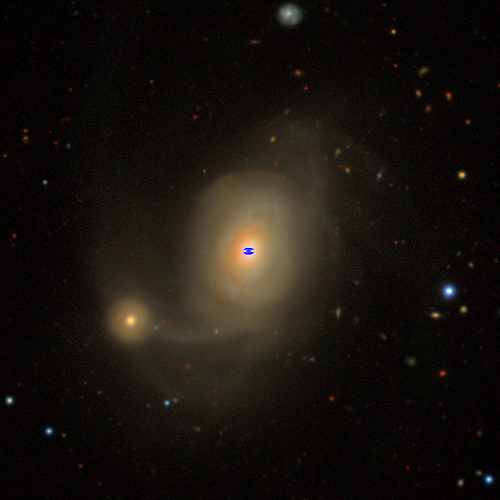}%
        }\hfill
        \subfloat[BAT 1189 (Ours, deep stretch)]{%
            \includegraphics[width=.32\linewidth]{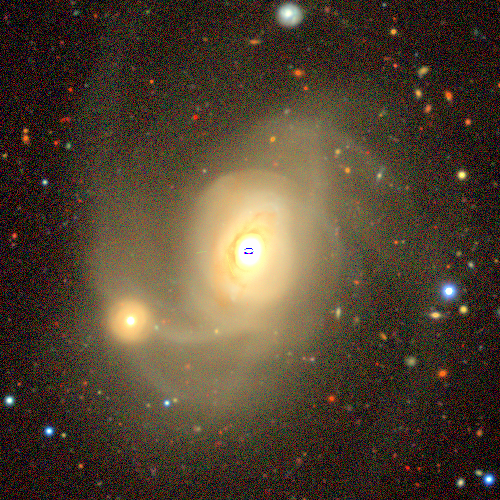}%
        }\\
        \subfloat[BAT 1180 (GZD)]{%
            \includegraphics[width=.32\linewidth]{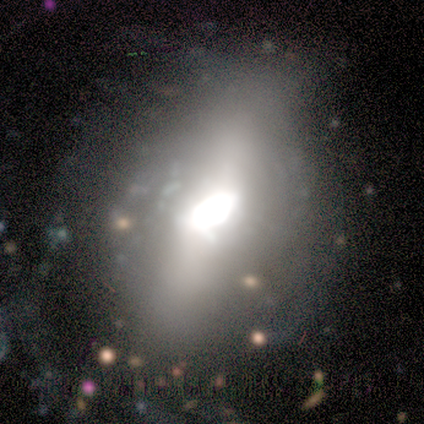}%            
        }\hfill
        \subfloat[BAT 1180 (Ours, shallow stretch)]{%
            \includegraphics[width=.32\linewidth]{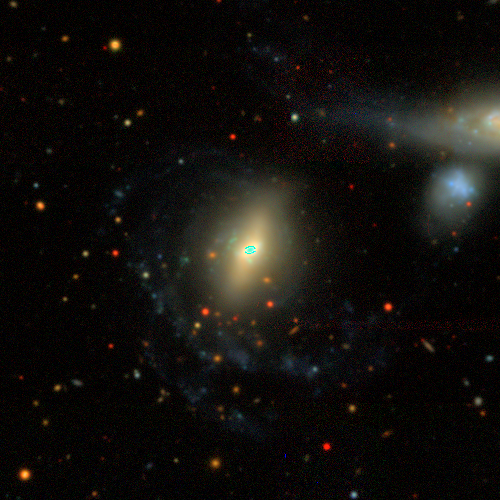}%
        }\hfill
        \subfloat[BAT 1180 (Ours, deep stretch)]{%
            \includegraphics[width=.32\linewidth]{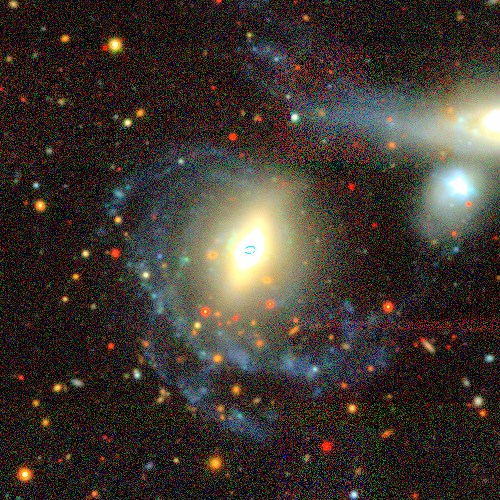}%
        }\\
        \subfloat[BAT 116 (GZD)]{%
            \includegraphics[width=.32\linewidth]{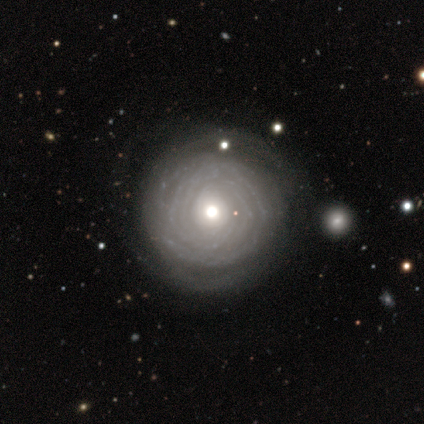}%
        }\hfill
        \subfloat[BAT 116 (Ours, shallow stretch)]{%
            \includegraphics[width=.32\linewidth]{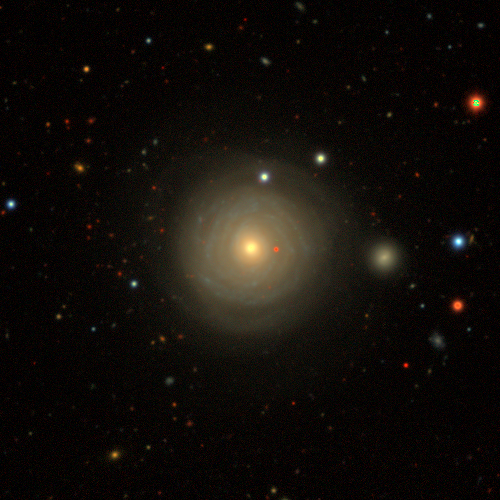}%
        }\hfill
        \subfloat[BAT 116 (Ours, deep stretch)]{%
            \includegraphics[width=.32\linewidth]{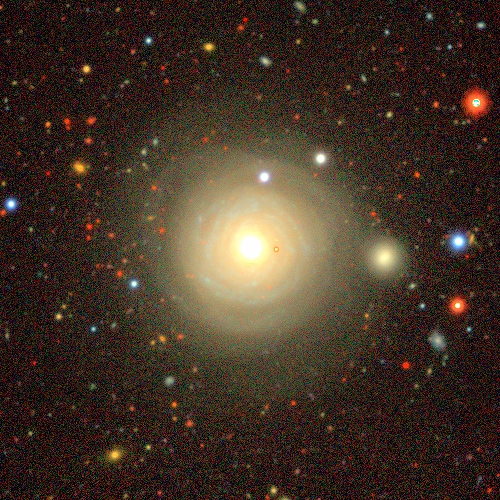}%
        }
        \caption{Selection of galaxies classified differently by GZD and our volunteers. Top row: Galaxy classified as other-uncertain by GZD (reclassified as disk Spiral) and as merger-strongly disturbed by us, most likely from the different stretches. Middle row: Galaxy classified as other-uncertain by GZD and merger-strongly disturbed by us, likely based on the larger FOV context and deeper stretch image. Bottom row: Galaxy classified as other-uncertain by GZD and disk Spiral by us, possibly because of the coloring.}
        \label{fig: triple-comparison}
\end{figure}

The most critical factor could arise from our decision to generate pairs of images with different stretches instead of a single one with a shallow stretch. This allowed us to highlight different cutout features and translated into volunteers making assessments with more information than they would have had otherwise, particularly about tidal tails, as shown in the top row of Fig. \ref{fig: triple-comparison}. With this sensitivity to mergers taken into account, one possibility is that objects that did not make the confidence cuts under GZD were classified as mergers by us, as evidenced by the confusion matrix in Fig.~\ref{fig:confusion}.

Another factor in identifying mergers, although to a lesser degree, can stem from the FOV we adopted for the cutouts. As mentioned in previous sections, we defined the FOV of our cutouts from redshift measurements and angular distance estimations, which were then fine-tuned accordingly. This resulted in imaged areas that, although similar to the Petrosian radius approach from GZD in most cases, could differ dramatically for some merger cases. Images from GZD were generally more zoomed in than ours, especially for mergers, to the point where an early merger stage companion galaxy or late merger stage tidal features might not be fully contained within the imaged area. This can be a deciding factor since the merger signatures can reside outside the cutouts' boundaries, as shown in the middle row of Fig. \ref{fig: triple-comparison}. 

Finally, although possibly less influential, there are differences in the coloring. While we decided to generate colorful images using the different bands available, the GZD project deliberately made the images washed out and uniform in color to lower the chance that their ML algorithm would incorporate that information into its classification process \citep{walmsley2020_colors}. This makes certain colorful features less obvious to humans, including spiral arms or dust lanes, which are relatively informative. The bottom row of Fig.\ref{fig: triple-comparison} shows a galaxy in which GZD volunteers did not agree on the presence of spiral arms, which become clearer in the colorful images.

In short, we expect BAT AGN to be successfully classified more frequently than GZD galaxies, as manifested by a lower relative fraction of other-uncertain for BASS. However, this tendency can be explained by the systematic differences between our study and the one conducted by GZD, specifically regarding the information in the color and stretches we provide to our volunteers. However, with so few volunteers looking at the BASS AGN, our classifications could suffer from overconfidence bias.

\subsection{Morphological class fractions}\label{sec:results_comp_gzd}

\begin{figure}
    \centering
    \includegraphics[width = \linewidth]{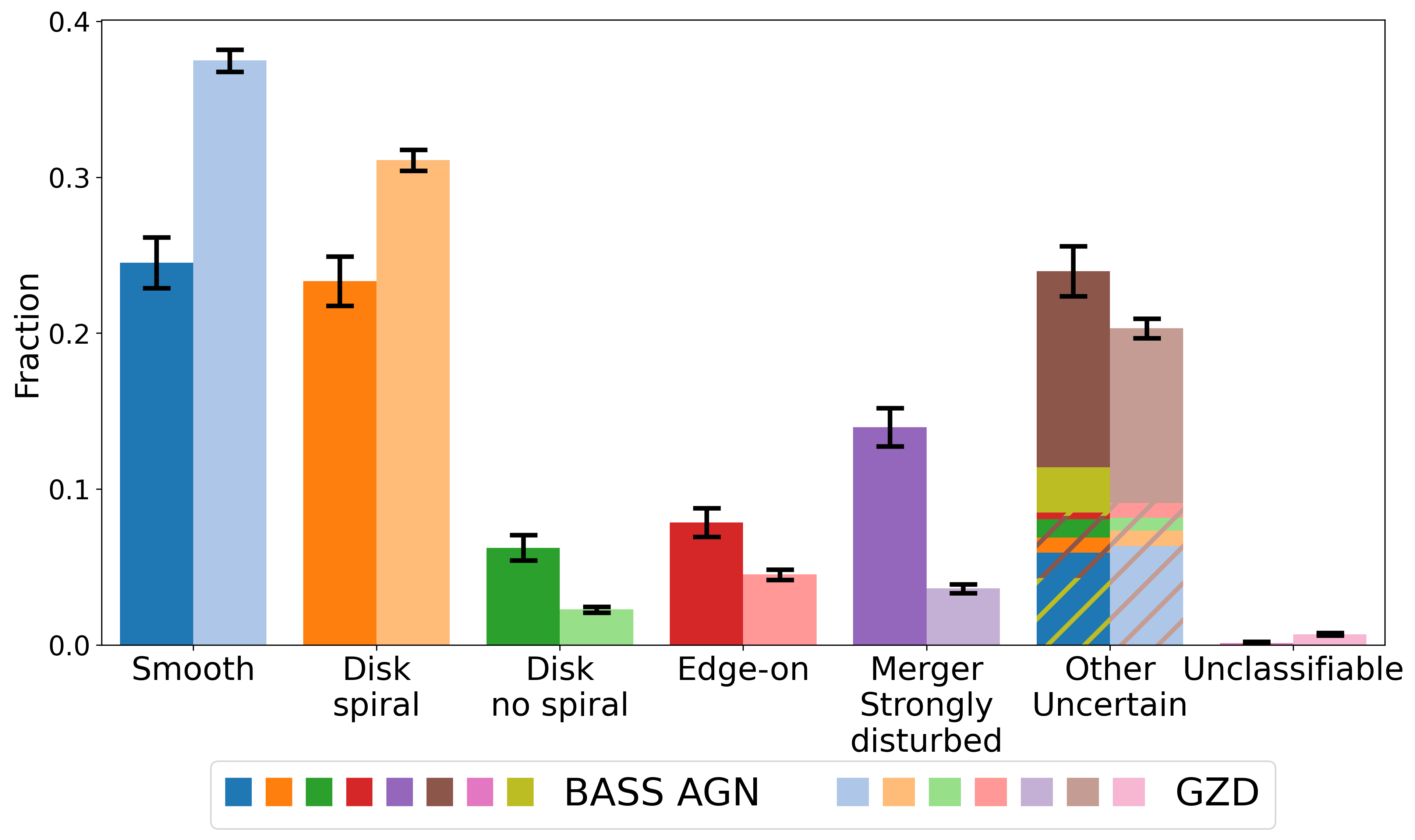}
    \caption{Comparison of the morphological class fractions between BASS AGNs and median comparable GZD samples. Error bars are 1$\sigma$ of Gauss statistics for BASS, and the distance from the median to the 16th and 84th percentile for GZD. Point-like AGNs are represented by the yellow bar stacked in the other-uncertain bar. Reclassified other-uncertain galaxies are shown as striped bars color-coded by their secondary morphology.}
    \label{fig: morpho comp}
\end{figure}

We compared the morphological class fractions among the comparable BASS sample and the median of 6000 GZD samples, which were generated from resamples that ensured that they had similar redshift and \textit{i}-mag distributions to the BASS sample. The comparison between their morphological class fractions is shown in Fig.~\ref{fig: morpho comp}. 

We find that the smooth class is visibly underrepresented in the BASS sample relative to GZD at $\sim6.4\sigma$ significance, or in other words, the BASS sample lacks smooth galaxies. These are galaxies associated with passive ellipticals, which are typically poor in cold gas that allows for star formation to take place; however, the same gas can also serve as fuel for the activation of the nucleus, as long as there is enough of a supply that allows for accretion at a rapid enough pace. Since smooth galaxies are poor in gas, activating the nucleus is an uphill battle for the galaxy in question, which can result in the relative scarcity of smooth galaxies among BASS AGN. This result is in line with the findings in \cite{Koss2011_BASSMorphs}, who also found a lack of elliptical galaxies among BASS AGN using Galaxy Zoo and RC3 \citep{RC3_1994AJ....108.2128C} morphologies.

Two of the three categories associated to disks are more prevalent in the BASS sample than the GZD sample, specifically the disk no spiral and edge-on by a significance of $\sim5.2\sigma$ and $\sim3.4\sigma$, respectively. The excess of spiral-less disks among AGN is particularly noteworthy, as these are typically associated with green valley galaxies slowing down their star formation. As the gas budget of spiral galaxies is internally depleted, their signature spiral arms start fading with the disk structure following suit \citep{Tinsley1980}. Considering that such galaxies could be undergoing a transitional phase of their evolution, the fact that the AGN sample contains a significantly higher amount of them could indirectly hint toward the role of AGN in suppressing their star formation. Moreover, the fact that the arms fade away while central AGN activity remains in place would suggest that the gas in the inner regions is yet to be depleted, since it is still fueling the AGN. This could indicate that the quenching is taking place inwards from the outskirts, a possibility showcased in some morphological transformation scenarios \citep[e.g.,][who reported a transition to outside-in growth for galaxies with $M_* < 10^{10} M_\odot$]{perez2013_outsidein_growth} colliding with inside-out quenching  \citep[e.g.,][]{Tacchella2015_insideout_quench}. However, extrapolating such claims based on this information alone could be dangerous, since it only follows from visual classifications and the role of AGN in the suppression of star formation remains openly debated in the literature \citep[e.g.,][]{Woodrum2024_agn_in_green_valley}. 

Perhaps noteworthy, BASS is markedly poor in spiral galaxies by $\sim4.3\sigma$, in direct opposition to its richness in the other two disk classes. Aggregating the family of disk categories into a single generic disk class bridges the gaps between the samples, with both quantities being consistent within $\sim0.1\sigma$.

The largest difference can be seen in the merger category, where the GZD fraction is $\sim8.4\sigma$ lower than the one calculated for BASS, or in terms of relative fractions, mergers and strongly disturbed galaxies are almost four times more frequent in the BASS sample than GZD. Mergers and nuclei activation have been linked for decades \citep{Barnes1991_ancient_link}, with gas inflows produced by the strong intergalactic interaction being ultimately responsible for the activation of the nucleus, and it is a possible interpretation of the relatively higher population of AGNs in merger systems \citep{treister2012_merger_lumagn, Ricci2017_Merger_growth}. Among these excess mergers in BASS AGN, we can identify some galaxies that could fit our definition for smooth and disk spiral galaxies (barring the low \texttt{Merging + Strongly disturbed} condition), but their contributions to the BASS AGN fractions are not enough to bridge the gap with GZD (extra 3.4\% for smooth and extra 3.3\% for disk spiral); thus our conclusions remain unaffected by them.

Finally, uncertain or unsuccessful classifications appear to have a similar incidence in both samples studied, falling within $\sim0.7\sigma$ in Figure \ref{fig: morpho comp}. Caution must be taken with this category, which, by definition, negates all the other categories. That being said, considering reclassifications alongside the other-uncertain class, the overall fractions get significantly lower because a considerable fraction of these galaxies are assigned a secondary morphology from the reclassification process, not to mention that we have defined a point-like class for BASS AGN, which has no proper equivalent in GZD. Regardless, including secondary morphologies is not enough to bridge the aforementioned gaps between classes of the disk family or the smooth galaxies, and comparing only the other-uncertains with no secondary morphologies assigned to them yields a mild excess for GZD at $\sim1.2\sigma$.

\begin{figure}[]
        \subfloat[BAT 138 - Unclear spiral-arms]{%
            \includegraphics[width=.48\linewidth]{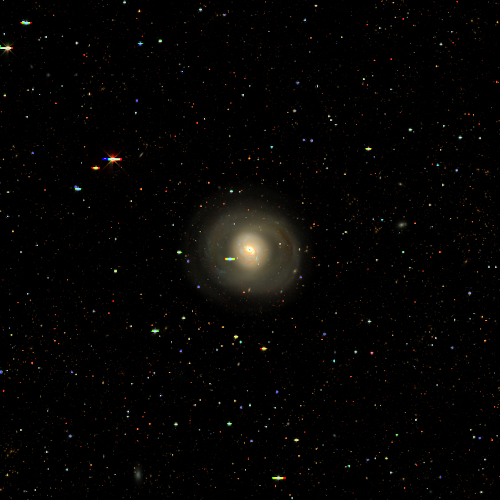}%
        }\hfill
        \subfloat[BAT 163 - Unclear disturbance degree]{%
            \includegraphics[width=.48\linewidth]{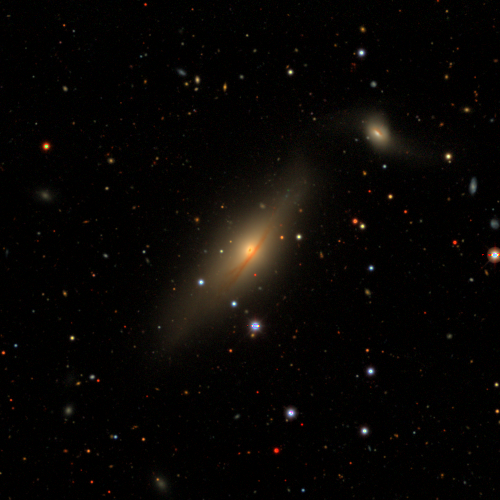}%
        }\\
        \subfloat[BAT 416 - Unclear broad morphology]{%
            \includegraphics[width=.48\linewidth]{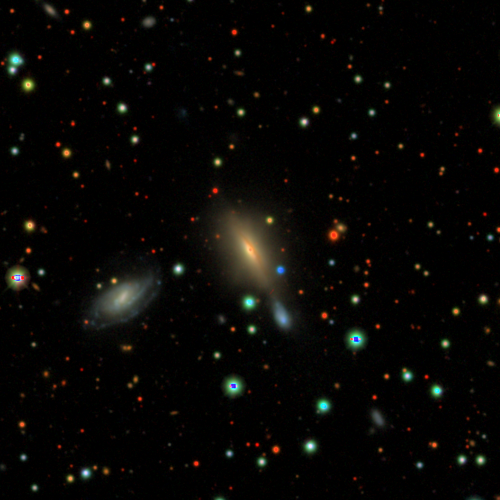}%
        }\hfill
        \subfloat[BAT 505 - Unclear disturbance degree]{%
            \includegraphics[width=.48\linewidth]{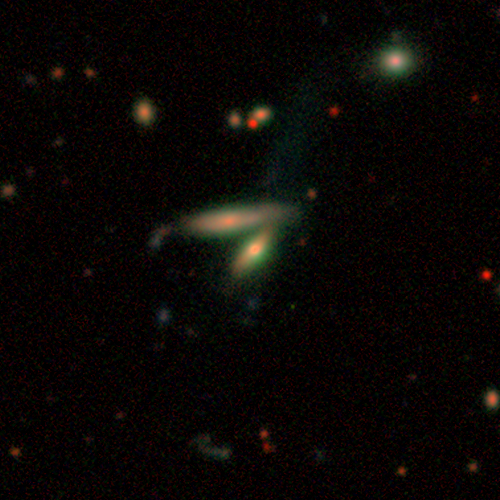}%
        }
        \caption{Various galaxies within the other-uncertain category, highlighting possible conflicts in interpretation.}
        \label{fig: uncertains}
    \end{figure}

Visual reinspection of the uncertain objects shows that the categories or traits in tension are mainly the following: disks with unclear spiral arms versus disks with no spiral arms, mergers that do not appear strongly disturbed (possibly still at an early stage) versus Weak disturbances, and smooth galaxies with dust lanes versus Lenticular galaxies. Examples of currently uncertain galaxies can be seen in Figure \ref{fig:compilation} (bottom row) as well as Fig.~\ref{fig: uncertains}, representing a range of possible interpretations.

The other-uncertain category has the most potential for variation as more classifications become available; as more opinions come in about these dubious objects, their status as a failed classification can either be obsolete or further cemented. Still, some galaxies are exotic in morphological terms.

In summary, we find that in relation to galaxies from GZD, BASS AGN dramatically prefer Merging/Strongly disturbed environments as well as spiral-less and edge-on disks to a lesser degree, while avoiding smooth and spiral galaxies. This seems to paint a picture in which gas budgets are key for the proliferation of nuclei activity. 

\subsection{Bar fraction}

Bar structures are thought to contribute to nuclear activity by funneling gas from the outskirts of the galaxy toward its center \citep{friedli1993_bar_gas_funnel}. The strength of the bars has also been found to positively correlate with nuclear activity, with strongly barred galaxies exhibiting a larger AGN fraction than both weakly barred and unbarred galaxies \citep[e.g.,][]{garland2024_bar_strength_agn}, although said relationship has also been put to scrutiny \citep[e.g.,][]{cisternas2013_no_link_bar_fuel}.

\begin{figure}
    \centering
    \includegraphics[width=0.95\linewidth]{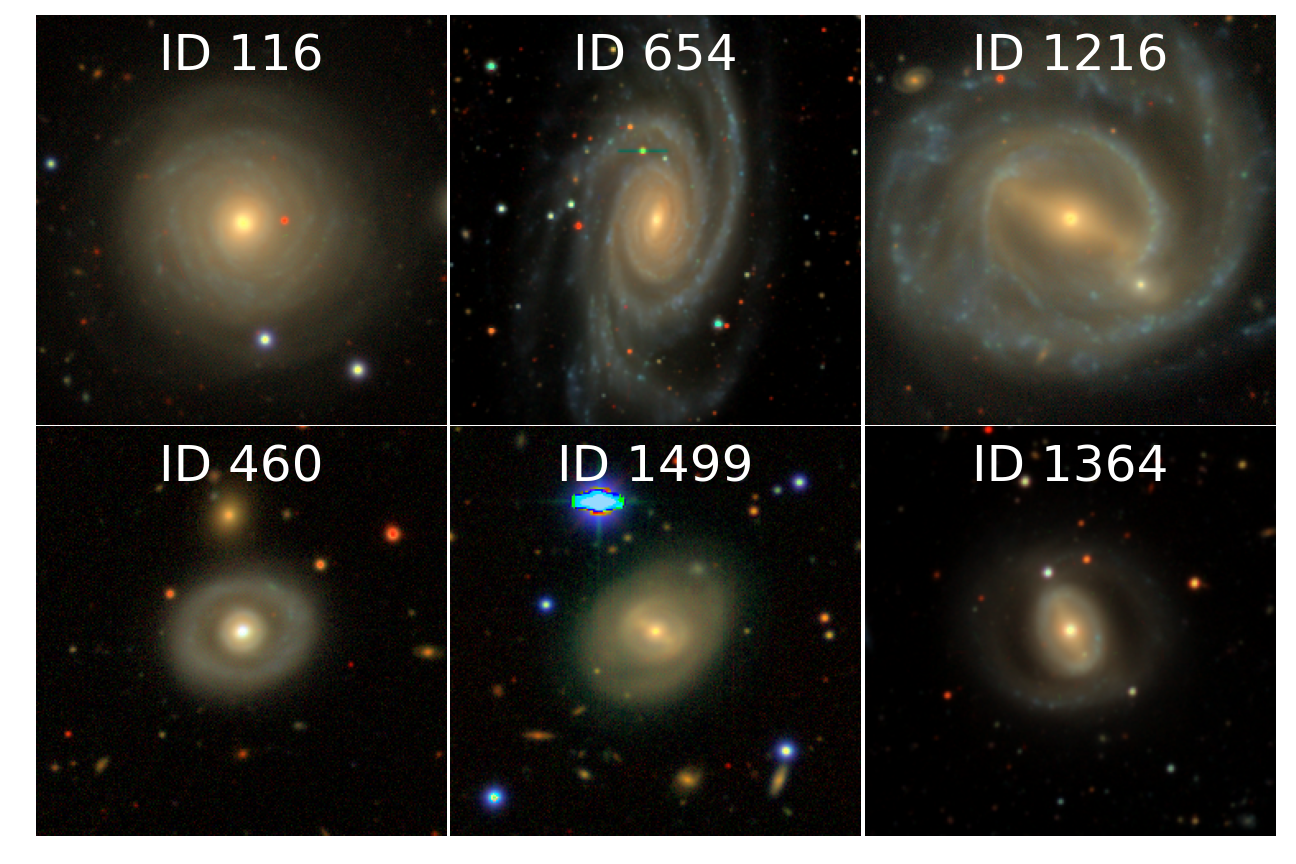}
    \caption{Examples of the different bar strengths among BASS AGNs for disk spiral (top) and disk no spiral (bottom). Left column: Unbarred. Center column: Weak bar. Right column: Strong bar. Images have been zoomed in to show the galactic centers better.}
    \label{fig:bar compilation}
\end{figure}

\begin{figure}
    \centering
    \includegraphics[width=0.95\linewidth]{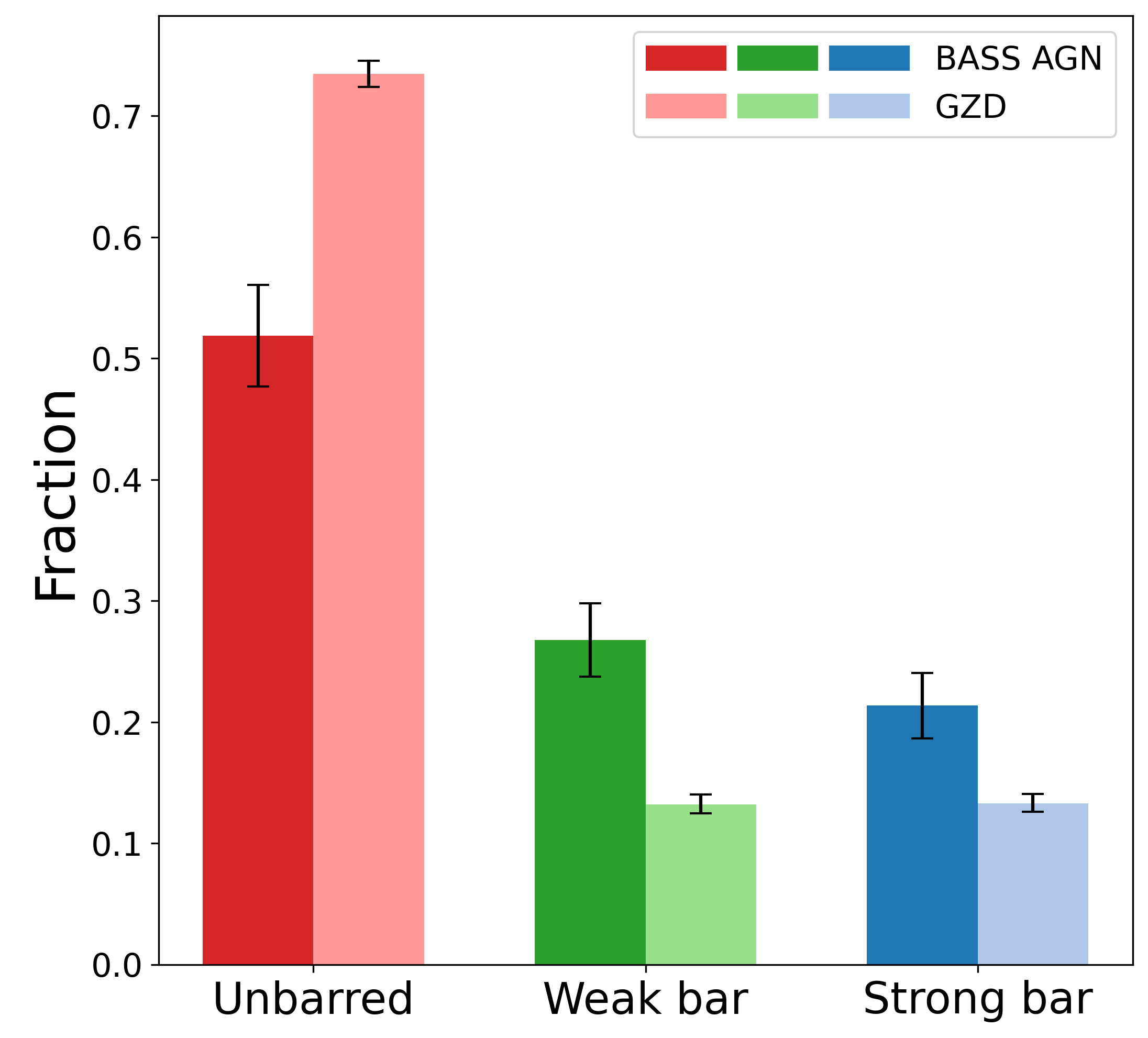}
    \caption{Comparison of the different degrees of bar strength between BASS AGNs and median comparable GZD samples. Error bars are 1$\sigma$ of Gauss statistics for BASS, and the distance from the median to the 16th and 84th percentile for GZD.}
    \label{fig:bar-strength}
\end{figure}

To investigate the incidence and strength of bars as a function of nuclei activity, we obtained the bar fraction of BASS AGN in face-on disks by defining subpopulations that are unbarred (\texttt{Bar no} $> 0.5$), weakly barred (\texttt{Bar strong} $+$ \texttt{Bar weak} $\geq 0.5$ and \texttt{Bar weak} $>$ \texttt{Bar strong}) and strongly barred (\texttt{Bar strong} $+$ \texttt{Bar weak} $\geq 0.5$ and \texttt{Bar weak} $\leq$ \texttt{Bar strong}), and compared said fractions to a matched (redshift and \textit{i}-mag) sample of inactive galaxies from GZD subjected to the same constraints.  Examples of the varying degrees of bar strength are shown in Figure \ref{fig:bar compilation}, whereas the comparison is shown in Figure \ref{fig:bar-strength} and the respective bar fractions and uncertainties are listed in Table \ref{tab:bar_fraccs}.

\begin{figure}
    \centering
    \includegraphics[width=0.95\linewidth]{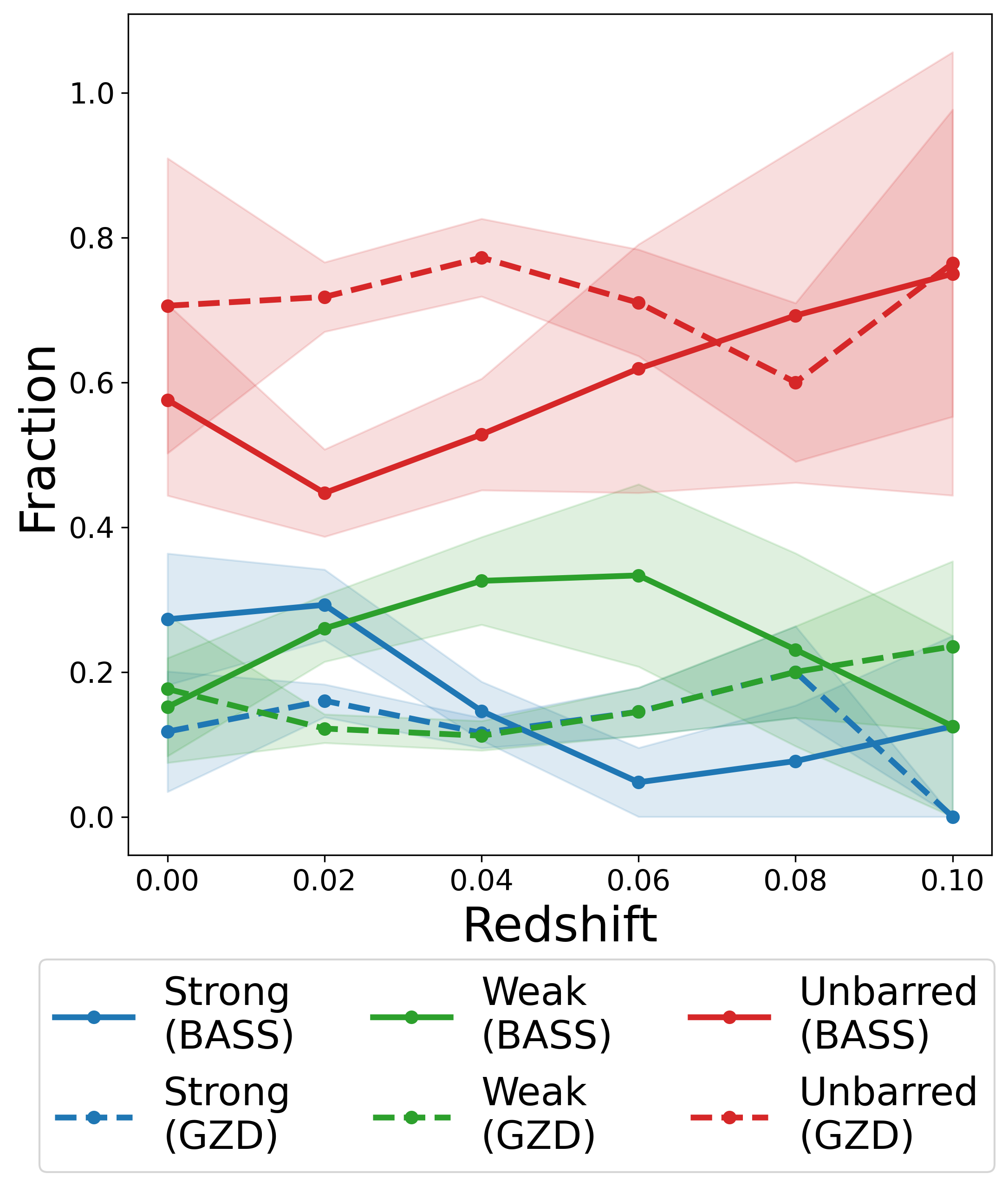}
    \caption{Evolution of the different degrees of bar strength as a function of redshift for BASS AGNs (continuous lines) and GZD (dashed lines). Shaded regions correspond to 1$\sigma$ significance for BASS, and the distance from the median to the 16th and 84th percentile for GZD.}
    \label{fig:bar-redshift}
\end{figure}

The computed fractions yield a significant excess of both weak and strong bars among BASS AGN with respect to inactive galaxies, with a difference of 4.3$\sigma$ and 2.9$\sigma$, respectively. Conversely, although a very small majority of the AGN sample is unbarred, they are relatively scarce compared to the inactive galaxies with a deficit of 5.0$\sigma$. These results are further supported when analyzing the bar fraction as a function of redshift, as shown in Figure \ref{fig:bar-redshift} where the BASS sample consistently exhibits a higher barred fraction than GZD in the better populated bins ($0 \leq z \leq 0.06$), beyond which the statistical effects become too dominant to draw meaningful conclusions. Regarding bar strengths, the weak and strong bar fractions are consistent within their uncertainties. 

\begin{table}[]
\centering
\caption{Percentage of bar strength category obtained for the active and inactive galaxies samples.}
\label{tab:bar_fraccs}
\begin{tabular}{llll}
\toprule
              & Unbarred                      & Weak bar                      & Strong bar                    \\ \midrule
BASS AGN      & 51.9 $\pm$ 4.2              & 26.8 $\pm$ 3.0              & 21.4 $\pm$ 2.7              \\
GZD inactives & 73.4$^{+ 1.0}_{- 1.1}$ & 13.2$^{+ 0.8}_{- 0.7}$ & 13.3$^{+ 0.7}_{- 0.7}$ \\ \midrule
\end{tabular}
\end{table}

\subsection{Bars and Eddington ratio}\label{sec:bar_edd}

\begin{figure}
    \centering
    \includegraphics[width=0.95\linewidth]{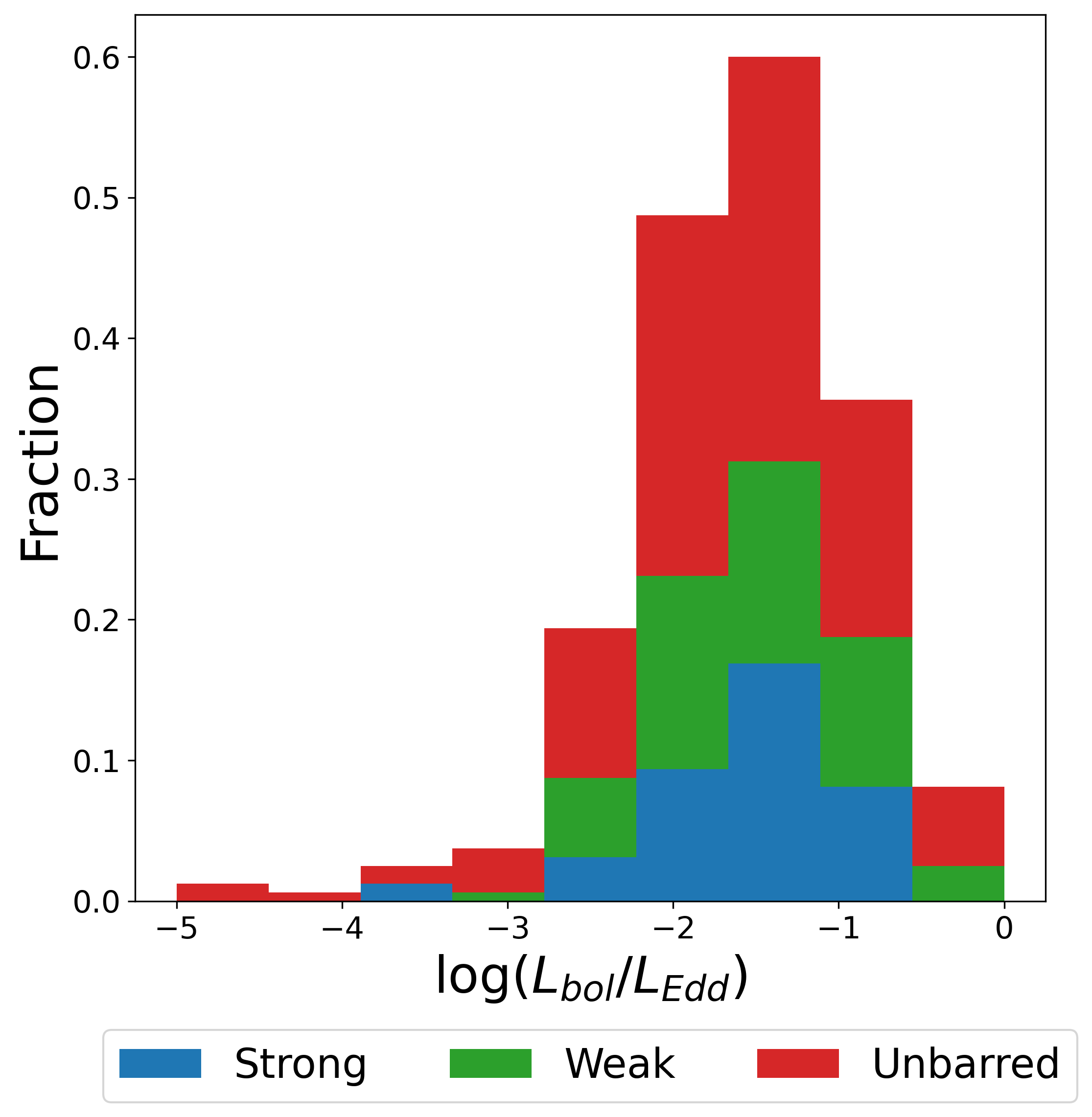}
    \caption{Stacked histogram of accretion rates for BASS AGNs according to their bar strength.}
    \label{fig:bar-edd_hist}
\end{figure}

\begin{figure}
    \centering
    \includegraphics[width=0.95\linewidth]{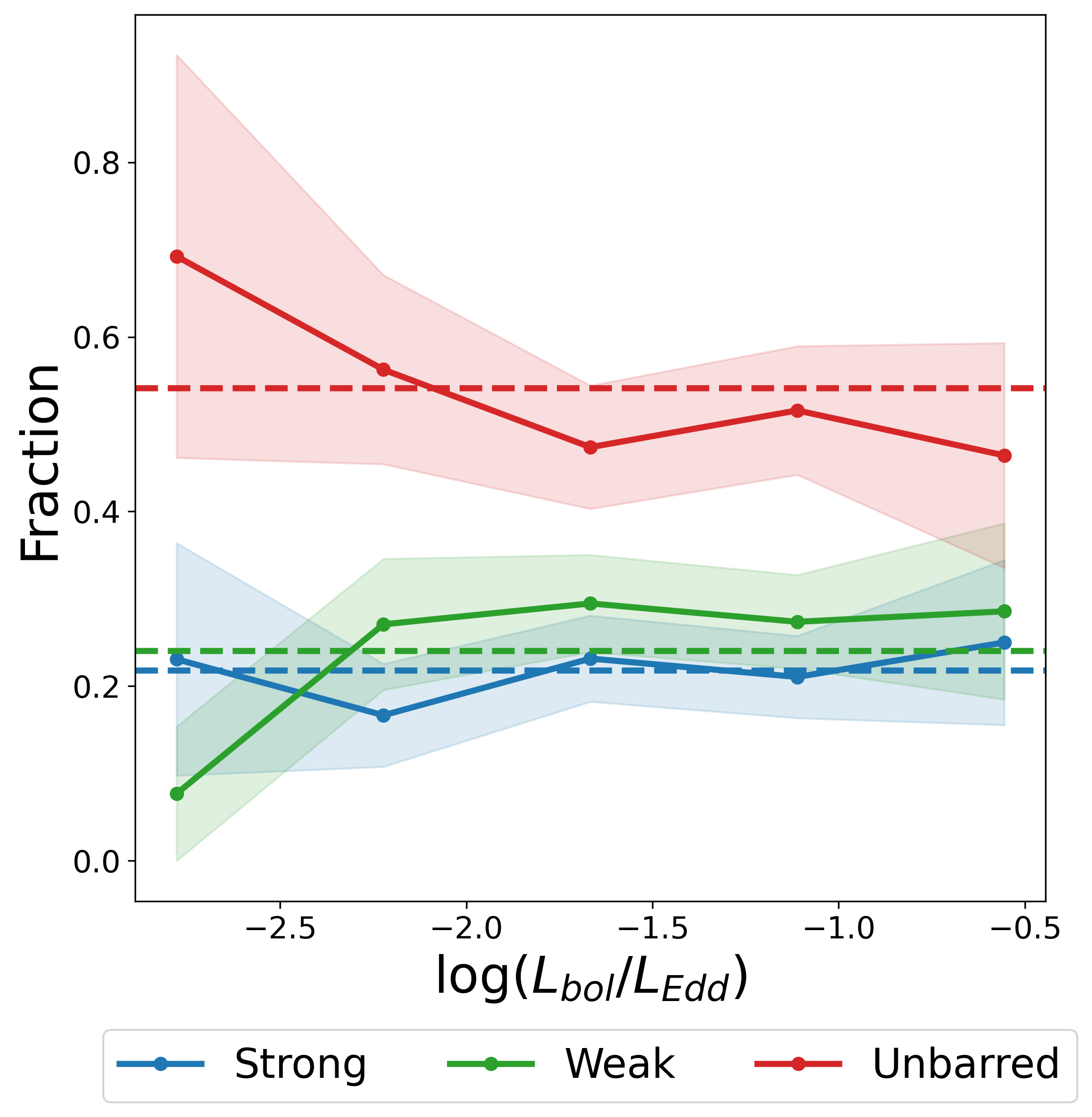}
    \caption{Evolution of the different degrees of bar strength as a function of accretion rate. Shaded regions correspond to 1$\sigma$ significance, while the dashed lines represent the mean value of each fraction. A small percentage of sources lie beyond the limits of the plot, but subsequently have poorly constrained bar fractions and were thus cropped out.}
    \label{fig:bar-trend}
\end{figure}

While the previous subsection focused on the bar fraction between active and inactive galaxies, we can also study the effects of bars on nuclei activity in terms of the accretion rate of the SMBH. Figure \ref{fig:bar-edd_hist} shows the distribution of accretion rates for the different degrees of bar strength, while Figure \ref{fig:bar-trend} compares the different bar fractions as a function of accretion rate.

The different classes of barred galaxies (strong, weak, unbarred) show roughly constant ratios between $-3{\lesssim}\log{(L_{\rm bol}/L_{\rm Edd})}{\lesssim}0$, as indicated in Figs.~\ref{fig:bar-edd_hist} and \ref{fig:bar-trend}, although there are too few sources at lower Eddington ratios to say anything conclusive about that regime. Despite the small tail of unbarred galaxies at lower accretion rates, a Kolmogorov-Smirnov test between barred and unbarred galaxies fails to reject the null hypothesis (Unbarred vs. Strong bar: $p$-value = 0.27, Unbarred vs. Weak bar: $p$-value = 0.39) and the populations are thus consistent with following the same distribution. If bars preferentially funnel gas toward the nucleus and positively influence the accretion rate of SMBHs, we might expect a trend between bar strength and accretion rate (e.g., unbarred AGN hosts populating the lower-end regime of accretion rates, strong bar hosts populating the upper regime). Figure~\ref{fig:bar-trend} indeed supports a marginal trend along these lines but suffers from limited statistics as only $\sim$300 galaxies are available for this bar analysis, where the trends are consistent with constant values within their uncertainties. Another caveat here relates to the lifetimes of bars and their efficiency in sustaining accretion, as both remain poorly constrained \citep[e.g.,][]{Bournaud2005,Ansar2025}, meaning that bars may dissipate before significant SMBH growth is observed.

\section{Summary and conclusions}\label{sec:conclusions}

The unbiased and complete nature of the BASS sample of AGNs with obscuration, which stems from their selection in hard X-rays, makes for a uniquely promising sample of galaxies with great scientific potential. We set ourselves to probe into the morphologies of AGNs, armed with said sample from BASS and a classification project mounted on Zooniverse. Once our volunteers had performed enough classifications on the galaxies, we defined the morphological classifications using their answers and attached general morphological labels to the objects in the sample. These newly acquired labels have considerable value on their own and can set the stage for a variety of scientific objectives, including, but not limited to, studying trends between AGN and host properties as a function of morphological classes and investigating whether or not AGNs show any preference in morphology when compared to a control sample of ``normal'' field galaxies from GZD. We intentionally based our classification scheme on that implemented during GZD-5 to facilitate a comparison with the GZD sample and gain a detailed overview of the sample. Still, our image coloration and stretches represent a key departure from the original study and have some implications for the observed morphologies.

With the data at hand, we recover well-known trends that link certain morphologies to several quantities related to the central SMBH, such as:

\begin{itemize} 
    \item Disk galaxies generally exhibit more modest values for their hard X-ray luminosity ($\log{(L_{\rm 14-195\,keV})} {\lesssim} 43.5$), black hole mass ($\log{(M_{\mathrm{BH}}/M_{\odot}) {\lesssim} 7.5}$), and Eddington ratio ($\log{(L_{\rm bol}/L_{\rm Edd})} {\lesssim} -2$).
    \item Point-like galaxies (and to a lesser degree smooth galaxies) generally correlate with higher values for hard X-ray luminosity ($\log{(L_{\rm 14-195\,keV})} \gtrsim 44$), black hole mass ($\log{(M_{\mathrm{BH}}/M_{\odot}) \gtrsim 8.5}$), and Eddington ratio ($\log{(L_{\rm bol}/L_{\rm Edd})} \gtrsim -1$).
\end{itemize}

When comparing active and inactive galaxy samples matched in redshift and \textit{i}-mag, we find that, relative to GZD:

\begin{itemize}
    \item The BASS AGN sample appears poor in smooth galaxies and rich in merging galaxies.
    \item The BASS AGN sample appears significantly poor in disk galaxies with spiral arms, yet rich in spiral-less disks and, to a lesser degree, edge-on galaxies. However, considering a generic disk class comprised of the above three classes, the BASS AGN sample has only a very mild deficit of disk galaxies.
    \item The BASS AGN sample exhibits a higher bar fraction ($\approx$48\% for BASS vs. $\approx$26\%), but we find no significant trend between bar strength and accretion rate.
\end{itemize}

Our results are generally consistent with the literature, wherein AGNs favor gas-rich environments with enough ``fuel'' to trigger activation, such as galaxy mergers. Recent Euclid results support this notion even at redshifts beyond $z = 0.5$, and find a higher merger fraction for AGNs across multiple selection criteria including X-ray \citep{euclid2025_merger_agn}. However, secular mechanisms such as bars remain an important path toward activation in our local sample. 

We tested for differences between obscured and unobscured systems using two definitions for obscuration. We found no evidence of morphological preferences relating to obscuration beyond an excess of edge-on galaxies on obscured AGNs or point-like galaxies in unobscured AGNs, so we concluded that obscuration does not significantly affect the results from \ref{sec:results_comp_gzd}. Regarding the classification success among BASS objects, we suspect that this has to do with the systematic differences in how we presented information to the volunteers rather than some intrinsic tendency of BAT AGNs toward clearer morphologies, specifically our colorful images and the usage of deep and shallow stretches.  

One of the objectives of BASS is to identify exotic populations of AGNs to gain more insight into them, including VP candidates. There are precedents for a search for VPs using optical imaging and volunteer classifications \citep[e.g.,][]{keel2012_vp_search} in a similar manner in which we gathered our data, and we identified a sample of VP candidates among BASS AGNs that we analyzed using MUSE IFU spectroscopy, in order to verify their gas clouds' ionization source. This sample of VP candidates will be introduced and released in a future paper (Parra Tello et al. in prep). In addition to VPs, the workflow we adopted provided us with information on traits and features that are yet to be explored for this sample, including, but not limited to, the presence of star-forming clumps, dust lanes, rings, or environmental effects from nearby companions not necessarily involving a merger. 

Finally, in addition to this project, it would be interesting to examine the morphologies of the hosts studied as part of the molecular gas analysis performed by \cite{Koss2021_BASS_Molgas}, to see if higher gas contents and AGN fueling could be gleaned (in a statistical fashion) from morphologies alone. Regarding atomic gas, a future paper has found evidence for group-like environments that significantly correlate with low-to-moderate luminosity AGN activity (Kim et al. in prep), a trend that could be contrasted or further supported with our data on nearby companions.

The morphological classifications presented in this work could be relevant for different analyses of the BAT AGN sample, specifically the various labels associated to objects and images. For starters, combining classifications and images makes for good ingredients for a training set if one is interested in designing an automatic classifier with ML based on Zoobot \citep{walmsley2022_zoobot}. More than 500 BASS AGN already have imaging from HST, which, combined with Euclid observations, will allow for the analysis of nuclear morphologies. The images and fits files themselves are also suitable for further study from a quantitative angle, such as performing morphological decomposition using GALFIT\footnote{\url{https://users.obs.carnegiescience.edu/peng/work/galfit/galfit.html}} \citep{peng2002_galfit, peng2010_galfitII} and/or estimating the bulge-to-total ratios.

\section*{Data availability}

Table \ref{tab: morpho_cat} is available in its entirety at the CDS via anonymous ftp to \texttt{cdsarc.u-strasbg.fr (130.79.128.5)} or via \texttt{http://cdsweb.u-strasbg.fr/cgi-bin/qcat?J/A+A/}.

\begin{acknowledgements}
We kindly thank the referee for the thorough reading of the manuscript and subsequent suggestions. We gratefully acknowledge funding from
ANID - Millennium Science Initiative - AIM23-0001 and ICN12\_009 (FEB), CATA-BASAL - FB210003 (FEB), and FONDECYT Regular - 1241005 (MPT, FEB). DD acknowledges PON R\&I 2021, CUP E65F21002880003, and Fondi di Ricerca di Ateneo (FRA), linea C, progetto TORNADO. CF acknowledges support from FONDECYT Postdoctoral grant - 3220751. IMC acknowledges support from FONDECYT Postdoctoral grant 3230653. AC and JK acknowledge support by the National Research Foundation of Korea (NRF), No. RS-2022-NR070872 and RS-2022-NR069020. KO acknowledges support from the Korea Astronomy and Space Science Institute under the R\&D program (Project No. 2025-1-831-01), supervised by the Korea AeroSpace Administration, and the National Research Foundation of Korea (NRF) grant funded by the Korea government (MSIT) (RS-2025-00553982).

This publication uses data generated via the Zooniverse.org platform, development of which is funded by generous support, including a Global Impact Award from Google, and by a grant from the Alfred P. Sloan Foundation.

The Legacy Surveys consist of three individual and complementary projects: the Dark Energy Camera Legacy Survey (DECaLS; Proposal ID \#2014B-0404; PIs: David Schlegel and Arjun Dey), the Beijing-Arizona Sky Survey (BASS; NOAO Prop. ID \#2015A-0801; PIs: Zhou Xu and Xiaohui Fan), and the Mayall z-band Legacy Survey (MzLS; Prop. ID \#2016A-0453; PI: Arjun Dey). DECaLS, BASS and MzLS together include data obtained, respectively, at the Blanco telescope, Cerro Tololo Inter-American Observatory, NSF’s NOIRLab; the Bok telescope, Steward Observatory, University of Arizona; and the Mayall telescope, Kitt Peak National Observatory, NOIRLab. Pipeline processing and analyses of the data were supported by NOIRLab and the Lawrence Berkeley National Laboratory (LBNL). The Legacy Surveys project is honored to be permitted to conduct astronomical research on Iolkam Du’ag (Kitt Peak), a mountain with particular significance to the Tohono O’odham Nation.

NOIRLab is operated by the Association of Universities for Research in Astronomy (AURA) under a cooperative agreement with the National Science Foundation. LBNL is managed by the Regents of the University of California under contract to the U.S. Department of Energy.

This project used data obtained with the Dark Energy Camera (DECam), which was constructed by the Dark Energy Survey (DES) collaboration. Funding for the DES Projects has been provided by the U.S. Department of Energy, the U.S. National Science Foundation, the Ministry of Science and Education of Spain, the Science and Technology Facilities Council of the United Kingdom, the Higher Education Funding Council for England, the National Center for Supercomputing Applications at the University of Illinois at Urbana-Champaign, the Kavli Institute of Cosmological Physics at the University of Chicago, Center for Cosmology and Astro-Particle Physics at the Ohio State University, the Mitchell Institute for Fundamental Physics and Astronomy at Texas A\&M University, Financiadora de Estudos e Projetos, Fundacao Carlos Chagas Filho de Amparo, Financiadora de Estudos e Projetos, Fundacao Carlos Chagas Filho de Amparo a Pesquisa do Estado do Rio de Janeiro, Conselho Nacional de Desenvolvimento Cientifico e Tecnologico and the Ministerio da Ciencia, Tecnologia e Inovacao, the Deutsche Forschungsgemeinschaft and the Collaborating Institutions in the Dark Energy Survey. The Collaborating Institutions are Argonne National Laboratory, the University of California at Santa Cruz, the University of Cambridge, Centro de Investigaciones Energeticas, Medioambientales y Tecnologicas-Madrid, the University of Chicago, University College London, the DES-Brazil Consortium, the University of Edinburgh, the Eidgenossische Technische Hochschule (ETH) Zurich, Fermi National Accelerator Laboratory, the University of Illinois at Urbana-Champaign, the Institut de Ciencies de l’Espai (IEEC/CSIC), the Institut de Fisica d’Altes Energies, Lawrence Berkeley National Laboratory, the Ludwig Maximilians Universitat Munchen and the associated Excellence Cluster Universe, the University of Michigan, NSF’s NOIRLab, the University of Nottingham, the Ohio State University, the University of Pennsylvania, the University of Portsmouth, SLAC National Accelerator Laboratory, Stanford University, the University of Sussex, and Texas A\&M University.

BASS is a key project of the Telescope Access Program (TAP), which has been funded by the National Astronomical Observatories of China, the Chinese Academy of Sciences (the Strategic Priority Research Program “The Emergence of Cosmological Structures” Grant \# XDB09000000), and the Special Fund for Astronomy from the Ministry of Finance. The BASS is also supported by the External Cooperation Program of Chinese Academy of Sciences (Grant \# 114A11KYSB20160057), and Chinese National Natural Science Foundation (Grant \# 12120101003, \# 11433005).

The Legacy Survey team makes use of data products from the Near-Earth Object Wide-field Infrared Survey Explorer (NEOWISE), which is a project of the Jet Propulsion Laboratory/California Institute of Technology. NEOWISE is funded by the National Aeronautics and Space Administration.

The Legacy Surveys imaging of the DESI footprint is supported by the Director, Office of Science, Office of High Energy Physics of the U.S. Department of Energy under Contract No. DE-AC02-05CH1123, by the National Energy Research Scientific Computing Center, a DOE Office of Science User Facility under the same contract; and by the U.S. National Science Foundation, Division of Astronomical Sciences under Contract No. AST-0950945 to NOAO.

The Pan-STARRS1 Surveys (PS1) and the PS1 public science archive have been made possible through contributions by the Institute for Astronomy, the University of Hawaii, the Pan-STARRS Project Office, the Max-Planck Society and its participating institutes, the Max Planck Institute for Astronomy, Heidelberg and the Max Planck Institute for Extraterrestrial Physics, Garching, The Johns Hopkins University, Durham University, the University of Edinburgh, the Queen's University Belfast, the Harvard-Smithsonian Center for Astrophysics, the Las Cumbres Observatory Global Telescope Network Incorporated, the National Central University of Taiwan, the Space Telescope Science Institute, the National Aeronautics and Space Administration under Grant No. NNX08AR22G issued through the Planetary Science Division of the NASA Science Mission Directorate, the National Science Foundation Grant No. AST-1238877, the University of Maryland, Eotvos Lorand University (ELTE), the Los Alamos National Laboratory, and the Gordon and Betty Moore Foundation.

This research uses services or data provided by the Astro Data Lab, which is part of the Community Science and Data Center (CSDC) Program of NSF NOIRLab. NOIRLab is operated by the Association of Universities for Research in Astronomy (AURA), Inc. under a cooperative agreement with the U.S. National Science Foundation.
\end{acknowledgements}

\bibliography{export-bibtex}
\bibliographystyle{aa}

\begin{appendix}
\section{Fringe pattern mitigation of Sinistro sources}\label{sec:Sinistro_fringe}

\begin{figure}[h!]
    \centering
    \includegraphics[width=\linewidth]{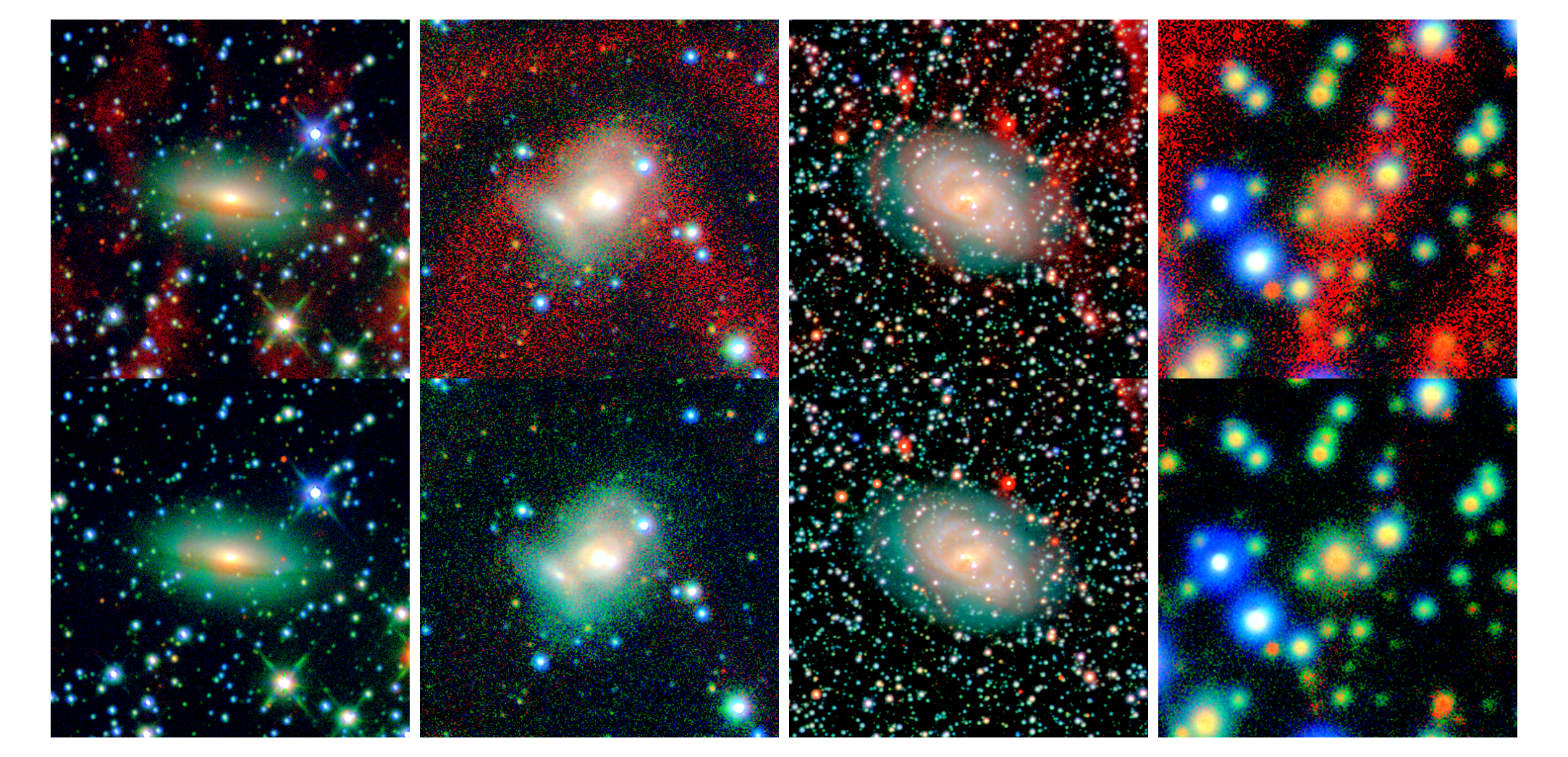}
    \caption{Examples of deep-stretch Sinistro images of BASS AGN host galaxies before (Top) and after (Bottom) mitigation of the fringing patterns.}
    \label{fig:fringo}
\end{figure}

Mitigation of fringing in Sinistro $z$-band images was achieved by clipping the values using a slightly higher floor than adopted for the other filters, and rescaling accordingly to cover the whole range of values. Since the intensity of the fringes was variable across the different observations, this floor was chosen on a case-by-case basis.
A comparison of galaxies before and after mitigating low-level $z$-band fringing from our observations is shown in Fig.~\ref{fig:fringo}, illustrating the different degrees of intensity of the fringes and the varying spatial resolutions they exhibited. 

\section{Dereddening of $A_{\rm V}\ge1$ sources}\label{sec:deredden}

\begin{figure}[h!]
    \centering
    \includegraphics[width=\linewidth]{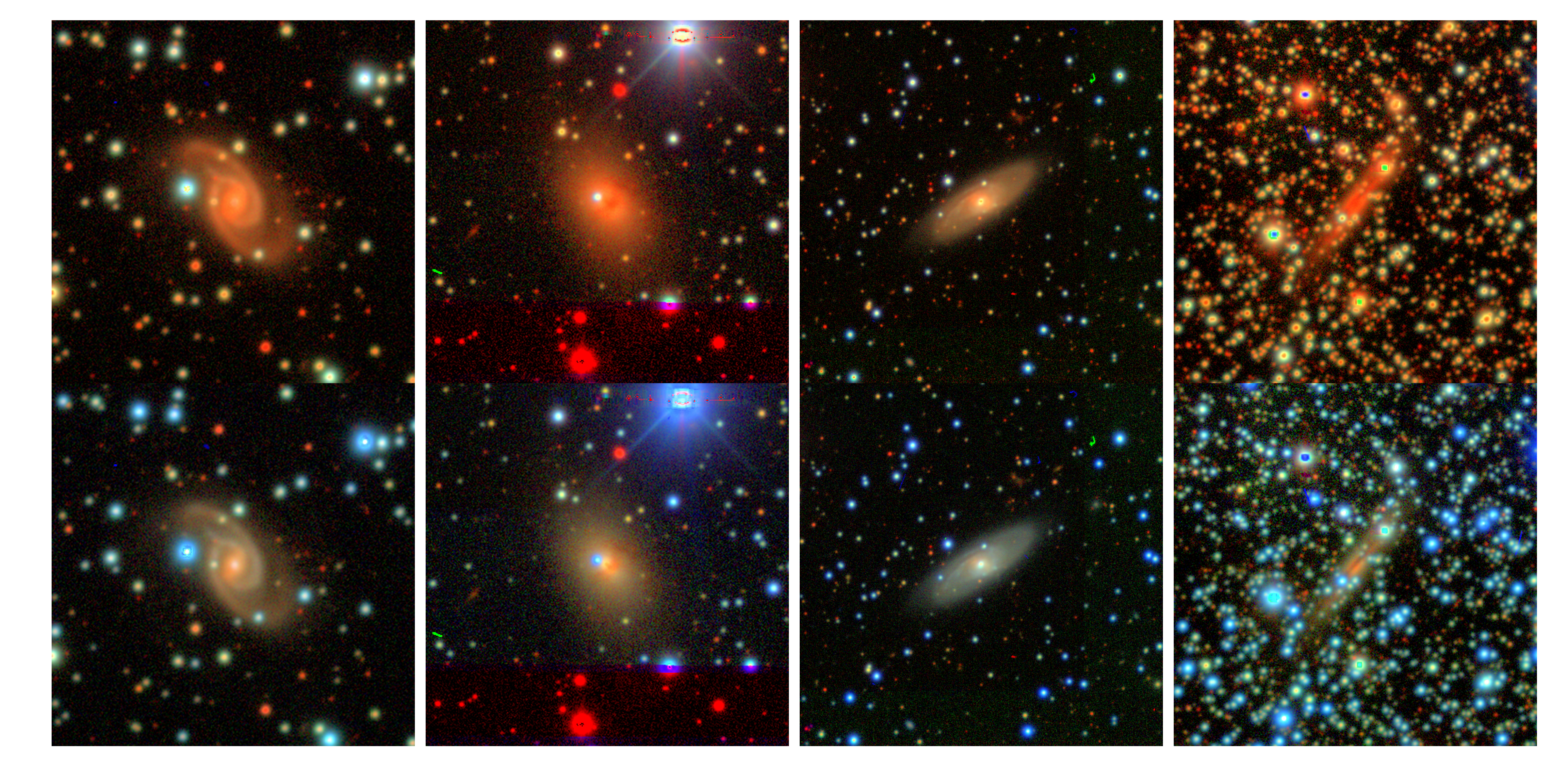}
    \caption{Selection of images of $A_{\rm V}\ge1$ galaxies before and after being subjected to MW de-reddening. Top: Before dereddening. Bottom: After dereddening.}
    \label{fig:dered}
\end{figure}

A comparison of galaxies before and after our de-reddening process is shown in Fig. \ref{fig:dered}.

\section{Redshift debias corrections} 
\label{sec:debias}

\begin{figure}[h!]
\centering
    \includegraphics[width = 0.9\linewidth]{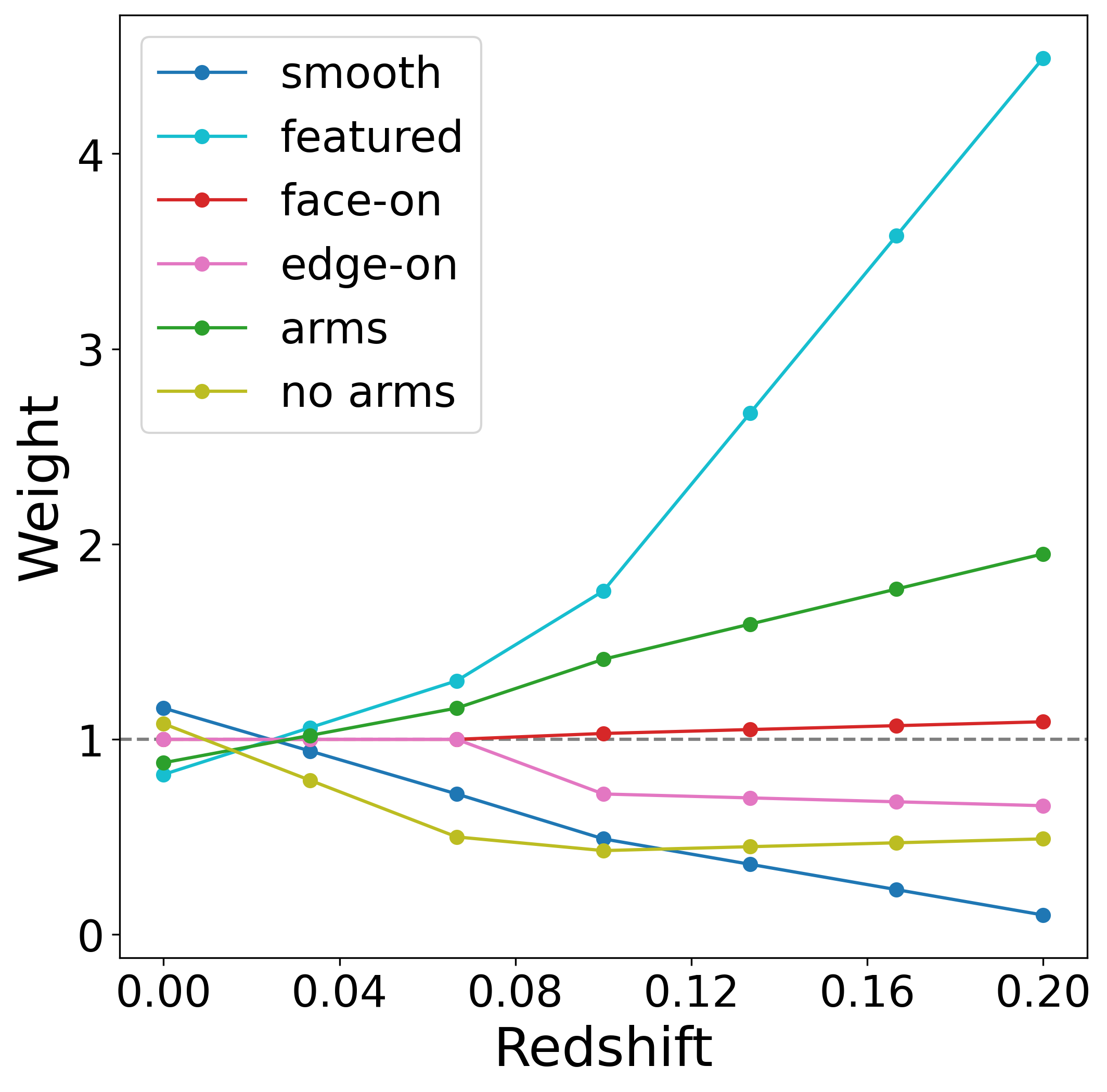}%
    \caption{Relative debias weights as a function of redshift for distinct morphological traits, adopted from \citet{Walmsley2022_gzd} and extrapolated to $z = 0$ and $z=0.2$.}
        \label{fig: trait_weights}
\end{figure}

Morphology assessment becomes harder and more strongly biased as relative spatial resolution and flux decrease with increasing redshift. For instance, disks with/without spiral arms or seen edge-on are harder to distinguish with redshift, while smooth galaxies appear strongly overestimated. To redshift-resolution debias the BASS AGN sample, we adopted the trends shown in Fig.~10 and discussed in Sec. 4.3.2 of \citet{Walmsley2022_gzd}. Therein, a relatively bright sample of GZD galaxies is adopted, which broadly overlaps with the redshift and magnitude parameter space of the BASS AGN sample. Specifically, we extract the trends for ``Smooth'', ``Featured'', ``Edge-on'', ``Face-on'', ``Spiral arms'' and ``No spiral arms'' and compute the ratio of the corrected and the original fractions, to obtain a correction weight for each of those traits as a function of redshift. We then extrapolate the trends to $z = 0$ and $z = 0.2$ using a first-order spline, which are shown in Fig.~\ref{fig: trait_weights}. It should be noted that the fractions in the original paper were presented with no errors associated with them.

\begin{figure}[h!]
\centering
    \includegraphics[width = 0.9\linewidth]{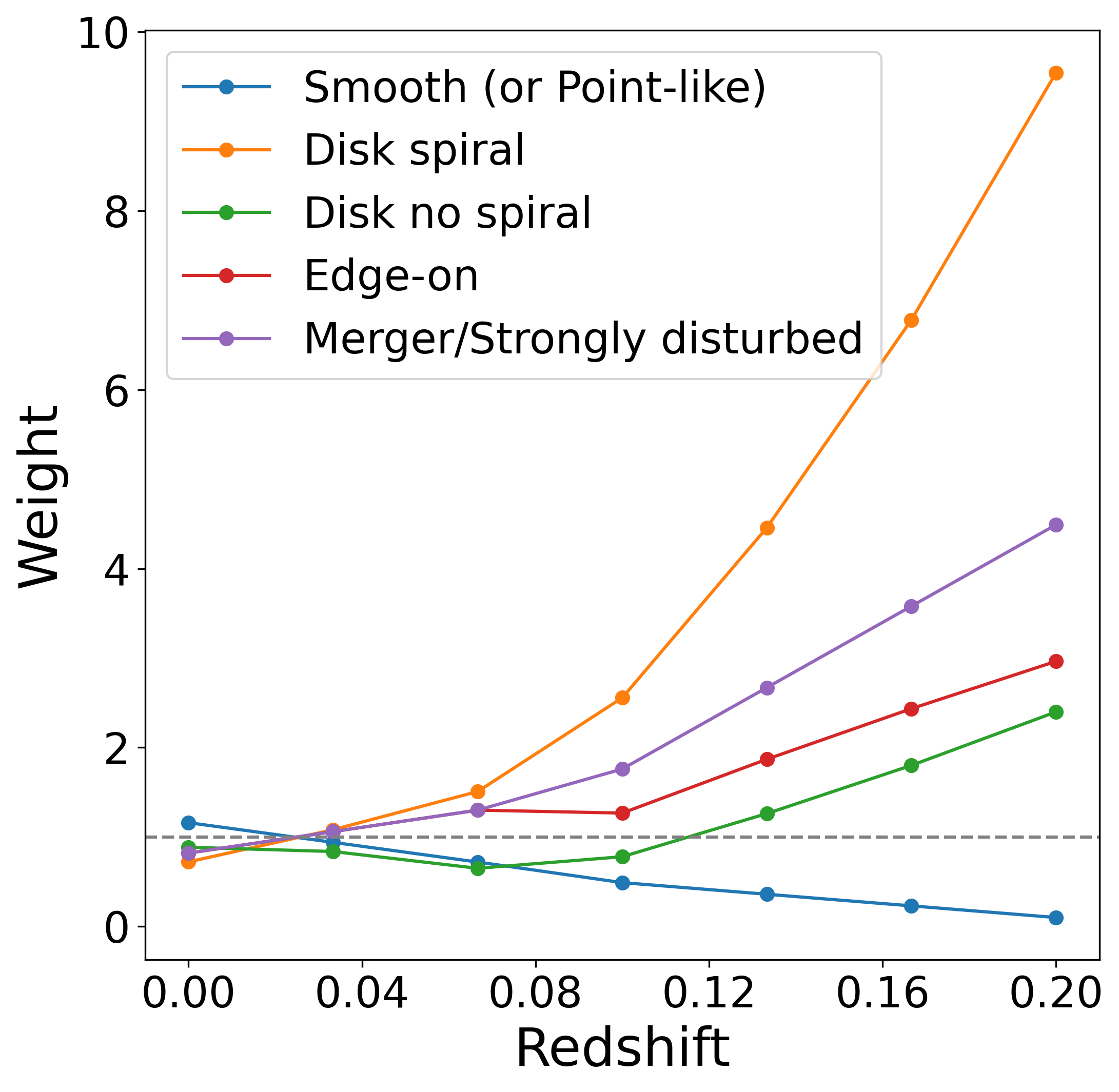}%
    \caption{Relative debias weights as a function of redshift for distinct morphological classes, adopted from \citet{Walmsley2022_gzd} and extrapolated to $z = 0$ and $z=0.2$.}
        \label{fig: class_weights}
\end{figure}

Since each of the classes depends on the chain of traits that they manifest (see Table \ref{tab: criteria}), we multiplicatively calculate a combined weight for the distinct morphologies based on the weights of the individual traits involved, such that the smooth class depends on the smooth correction, disk-spiral depends on the featured, face-on and spiral corrections, and so on. The point-like and merger-strongly disturbed classes required a workaround for defining their weights, since the former is an original inclusion of our workflow and the latter's corrections were impossible to extract. Thus, we assigned the smooth weight to the point-likes and the featured weight to the mergers, resulting in the corrections shown in Fig.~\ref{fig: class_weights}. Subsequently, when counting or assessing fractions, each source is weighted according to its redshift as interpolated from the Fig.~\ref{fig: class_weights} curves.

\section{On morphology and obscuration} 
\label{sec:results_BASS_NH}

\begin{figure}[h!]
        \subfloat[Seyfert selection]{%
            \includegraphics[width = \linewidth]{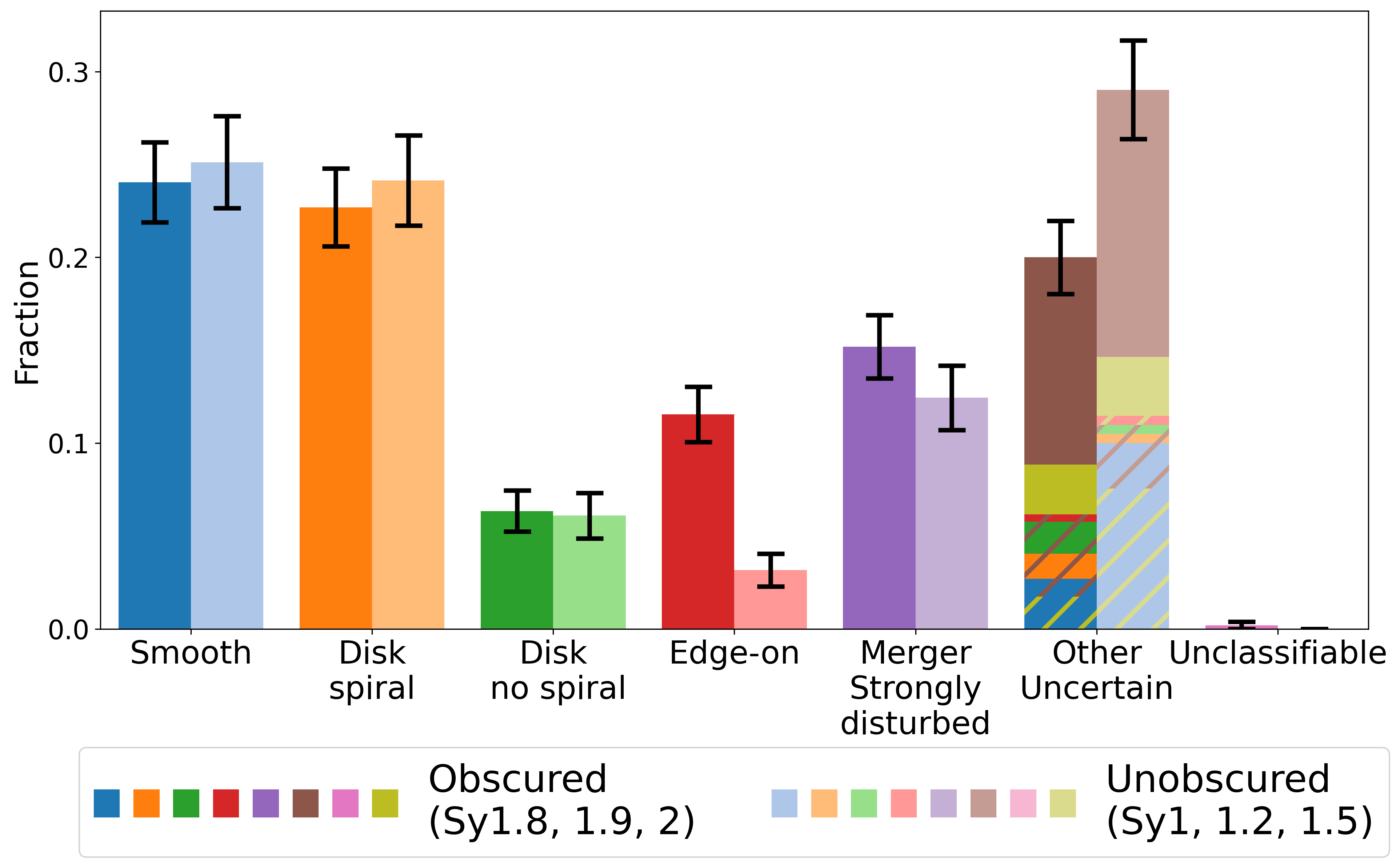}
        }\\
        \subfloat[Column density selection]{%
            \includegraphics[width = \linewidth]{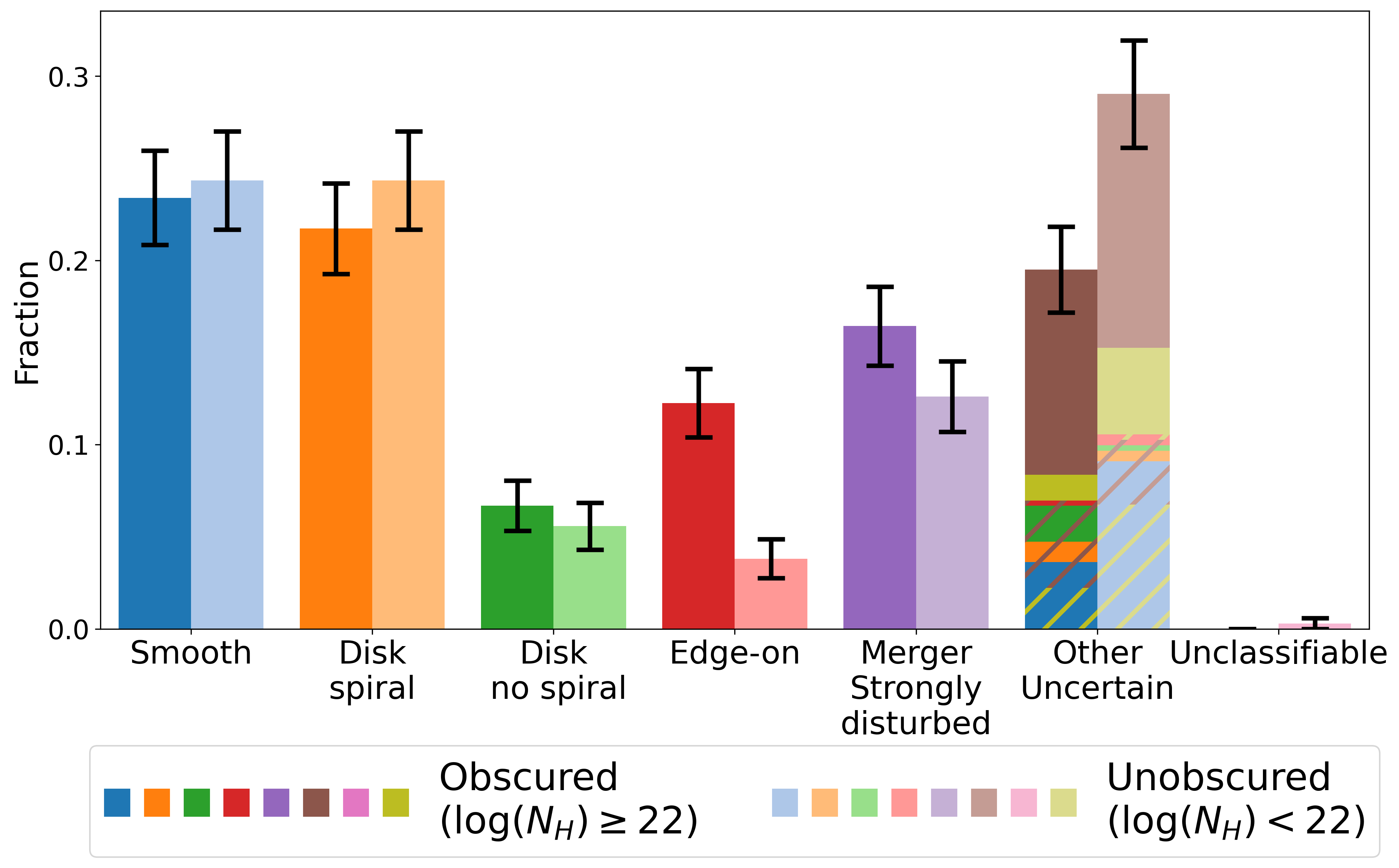}%
        }
        \caption{Morphological class fraction comparison between obscured and unobscured BAT AGN, separated by Seyfert class (top) and by hydrogen column density (bottom). Error bars are equivalent to 1$\sigma$ of Poisson errors.}
        \label{fig: (un)obscured}
\end{figure}

When we initially matched the magnitude distributions of BASS and GZD, we found that type 1/unobscured AGN tend to contaminate the host magnitude measurements far too much to allow for a reasonable match. Thus, we adopted a workaround by limiting the matching sample to obscured BASS AGN, given that they do not suffer from this problem. However, by doing so we implicitly assume that galaxies hosting obscured and unobscured AGN are not too different regarding morphology. To test this assumption, we compared the morphological class fraction between galaxies hosting obscured and unobscured AGN based on Seyfert type, which can be seen in the top panel of Fig.~\ref{fig: (un)obscured}. These galaxies comprised 930 sources with $z \le 0.15$ (520 obscured and 410 unobscured objects). 

After accounting for the error bars, the differences are not so noticeable, with most classes being separated by less than 1$\sigma$. Mergers seem to mildly favor obscured systems over unobscured ones, which could be a reasonable find considering that the gas inflows from mergers bring material toward the galactic center, which could then partake in the obscuration of the nucleus and the observed lack of broad lines; conversely, it would be slightly more unlikely for the nucleus to remain unobscured in such a chaotic scenario. Still, this $\sim1.12\sigma$ preference for obscuration and mergers is insufficient to consider some intrinsic difference between the samples. We also find no evidence of unobscured AGN favoring smooth hosts in our sample, which is not in line with literature results suggesting otherwise \citep[e.g.,][]{bornancini2018_agn1-2_ellip, gkini2021_sy1_sy2_morphos}. 

The most statistically significant remark is that despite disk spiral, disk no spiral, and edge-on being subcategories of the family of disk galaxies, there appears to be a preference of obscured galaxies toward the edge-on class at $\sim4.8\sigma$ significance, but no apparent preference for the other two disk subcategories. edge-on galaxies have long been known to harbor obscured AGN more often than face-on galaxies \citep[e.g.,][]{Maiolino1995_edge/face_obscuration, gkini2021_sy1_sy2_morphos}, and considering that this class is merely a denomination derived from the circumstantial inclination of the disk, this excess in edge-on galaxies could be explained by the contribution of the galactic plane to the overall obscuration of the nucleus, which is maximized under the edge-on configuration. 

Finally, there appears to be a 3.4$\sigma$ excess in other-uncertains for unobscured galaxies, driven primarily by the inclusion of point-likes in the stack; excluding this morphology from the calculations lowers the difference to 1.3$\sigma$. This relative prominence of point-like sources among unobscured AGN is expected, as it corresponds to a configuration whereby the AGN emission dominates over the host galaxy, facilitated by the lack of material obstructing the central SMBH. 

In parallel, we use available $N_\mathrm{H}$ measurements from \cite{Ricci2017_BAT_X-ray}, considering the same redshift range as before, as an alternate selection method for obscured AGN. We consider AGN with column densities of log$(N_\mathrm{H}) \geq 22$ to be obscured. Among 700 BASS AGN with $N_\mathrm{H}$ measurements, 359 are thus obscured and 341 unobscured. Although the split between both classes is more balanced than the previous approach, the availability of $N_\mathrm{H}$ measurements is more limited. The relative morphological classes under this alternate definition are shown on the lower panel of Fig.~\ref{fig: (un)obscured}, and appear virtually identical to the previous Seyfert class definition. Implementing a stricter threshold of log$(N_\mathrm{H}) \geq 23$ for the obscured galaxies does not affect our conclusions.

We conclude that, outside the expected irregularity regarding the excess of obscured AGN in edge-on galaxies, obscured and unobscured galaxies are fully consistent in terms of morphology whether we define them based on their broad lines (or lack thereof) or their column densities. This should maintain the integrity of the comparison with GZD from \ref{sec:results_comp_gzd} and its interpretation, which proceeded under this assumption.

\end{appendix}

\end{document}